\documentclass{aastex7}

\usepackage{booktabs}
\usepackage{CJK}
\usepackage{makecell}
\usepackage{amsmath}
\usepackage{caption}
\usepackage{xcolor}
\usepackage{multirow}
\usepackage{tabularx}
\usepackage{subcaption}

\newcolumntype{Y}{>{\centering\arraybackslash}X}
\begin{document}
\journalinfo{}

\begin{CJK*}{UTF8}{gbsn}

\title{A Cross-Laboratory Comparison Study of Titan Haze Analogs: Surface Energy II}

\correspondingauthor{Xinting Yu}
\email{xinting.yu@utsa.edu}

\author[0000-0002-9075-0723, sname=Austin, gname=Eric]{Eric C. Austin}
\affiliation{Department of Physics and Astronomy, University of Texas at San Antonio,
1 UTSA Circle, San Antonio, TX 78249, USA}
\email{eric.austin@utsa.edu}

\author[0000-0002-7479-1437, sname=Yu, gname=Xinting]{Xinting Yu (余馨婷)}
\affiliation{Department of Physics and Astronomy, University of Texas at San Antonio,
1 UTSA Circle, San Antonio, TX 78249, USA}
\email{xinting.yu@utsa.edu}

\author[0000-0002-6694-0965, sname=He, gname=Chao]{Chao He} % made 
\affiliation{Department of Earth and Planetary Sciences, Johns Hopkins University,
3400 N. Charles Street, Baltimore, MD, 21218, USA}
\email{che13@jhu.edu}

\author[0009-0007-6910-6347, sname=Pesciotta, gname=Cara]{Cara Pesciotta} % made 
\affiliation{Department of Earth and Planetary Sciences, Johns Hopkins University,
3400 N. Charles Street, Baltimore, MD, 21218, USA}
\email{cpescio1@jhu.edu}

\author[0000-0002-1883-552X, sname=Sciamma-O'Brien, gname=Ella]{Ella Sciamma-O'Brien} % made 4
\affiliation{Space Science and Astrobiology Division, NASA Ames Research Center, Astrophysics Branch, Moffett Field, CA, 94035, USA}
\email{ella.m.sciammaobrien@nasa.gov}

\author[0000-0001-9612-8532, sname=Sebree, gname=Joshua]{Joshua A. Sebree} % made 2
\affiliation{Department of Chemistry and Biochemistry, University of Northern Iowa,
1227 W. 27th Street, Cedar Falls, IA, 50614, USA}
\email{joshua.sebree@uni.edu}

\author[0000-0002-3554-5321, sname=Montes Bojorquez, gname=Jose Raul]{Jose Raul Montes-Bojorquez}
\affiliation{Department of Physics and Astronomy, University of Texas at San Antonio,
1 UTSA Circle, San Antonio, TX 78249, USA}
\email{joseraul.montesbojorquez2@utsa.edu}

\author[0009-0003-9055-7397, sname=Husic, gname=Adis]{Adis Husi\'c}
\affiliation{Department of Physics and Astronomy, University of Texas at San Antonio,
1 UTSA Circle, San Antonio, TX 78249, USA}
\email{adis.husic@utsa.edu}

\author{Erik White}
\affiliation{Department of Earth and Planetary Sciences, University of California Santa Cruz,
1156 High St., Santa Cruz, CA, 95064, USA}
\email{erscwhite@gmail.com}

\author[0009-0000-8181-3705, sname=Bond, gname=Christopher]{Christopher R. Bond}
\affiliation{Department of Materials Science and Engineering, Johns Hopkins University, 3400 N. Charles Street, Baltimore, MD, 21218, USA}
\email{cbond15@jhu.edu}

\author[0000-0003-4596-0702, sname=Horst, gname=Sarah]{Sarah M. H\"orst}
\affiliation{Department of Earth and Planetary Sciences, Johns Hopkins University,
3400 N. Charles Street, Baltimore, MD, 21218, USA}
\email{sarah.horst@jhu.edu}

\author[0000-0002-6064-4401, sname=Salama, gname=Farid]{Farid Salama}
\affiliation{Space Science and Astrobiology Division, NASA Ames Research Center, Astrophysics Branch,
Moffett Field, CA, 94035, USA}
\email{farid.salama@nasa.gov}

\author[0000-0003-3001-3163, sname=McGuiggan, gname=Patricia]{Patricia McGuiggan}
\affiliation{Department of Materials Science and Engineering, Johns Hopkins University,
3400 N. Charles Street,
Baltimore, MD, 21218, USA}
\email{mcguiggan@jhu.edu}

\keywords{Titan (2186) --- Saturnian Satellites (1427) --- Laboratory Astrophysics (2004) --- Experimental Data (2371) --- Experimental Techniques (2078) --- Planetary Atmospheres (1244) --- Planetary Science (1255) --- Planetary Surfaces (2113)}

\begin{abstract}
The organic haze particles on Titan play important roles in atmospheric cloud formation and aerosol-lake interactions. These processes are strongly influenced by the surface energy of the haze particles, which controls cohesion and wetting behavior. This study presents a comparative analysis of 32 laboratory-produced haze analog samples (``tholins'') synthesized across three laboratories. Using contact angle measurements, we determine the total surface energy and its dispersive and polar components for all samples, systematically evaluating the effects of substrate choice, air exposure, initial N$_2$/CH$_4$ gas mixture, and experimental setup. We find that tholin samples exhibit minimal substrate dependence, whereas exposure to ambient air substantially modifies the surface chemistry, altering the balance between dispersive and polar components. Thus, future Titan-relevant surface property measurements may use any substrate but must keep samples pristine. Surface energies vary weakly across methane concentrations, from 1--10\% CH$_4$ in N$_2$, indicating that Titan's hazes formed at different altitudes should exhibit broadly similar cohesiveness. In contrast, experimental conditions such as gas exposure time and energy source produce the dominant differences in surface energy, driven largely by variations in polar components. Despite these differences, tholin samples exhibit high dispersive components, implying that Titan's hazes should act as efficient cloud condensation nuclei for hydrocarbon clouds and should generally sink into Titan's lakes. Given observed ethane ice clouds, we conclude that cold plasma tholins may be better physical analogs for Titan's hazes than samples produced with far-ultraviolet irradiation, though intrinsic surface energies of UV tholins remain uncertain due to film thickness limitations.
\end{abstract}

\section{Introduction}
Titan, Saturn's largest moon, is enveloped in substantial haze layers composed of particles generated from the interaction of solar UV photons and energetic particles from Saturn's magnetosphere with its predominantly nitrogen (N$_2$) and methane (CH$_4$) atmosphere \citep{Niemann2005, Flasar2005, Horst2017, Nixon2024-ji}. This photochemically active environment drives the formation of increasingly complex organic molecules that further react and aggregate to form solid haze particles \citep{Khare2001-zm, Imanaka2004-nt, Waite2005, Zhao2018, Mouzay2021-bm, Nixon2024-ji}. These haze particles can serve as cloud condensation nuclei (CCN), possibly contributing to the observed cloud formations within Titan's dense atmosphere \citep{Barth2006, Curtis2008, Lavvas2011, Horst2017, Anderson+2018, Rannou2019, Yu2020, Li2022}. Upon sedimentation onto the moon's surface, they may also participate in aeolian and fluvial processes, potentially influencing the formation of dunes, the appearance and disappearance of `magic islands' on the hydrocarbon lakes, the dampening of waves, and the composition of lake sediments  \citep{Soderblom2007, Stephan2010, Barnes2011, Soderblom2012, Zebker2014, Grima2017, Lopes2019, Yu2020, Li2022, Yu2024}.

In laboratory settings, Titan haze analogs, commonly referred to as ``tholins", are synthesized using a range of gas mixtures and energy sources to simulate Titan's atmospheric chemistry under different conditions. These experiments typically expose N$_2$/CH$_4$ mixtures with trace amounts of other gases to energy sources such as UV lamps, synchrotron sources, femtosecond lasers, and cold plasma discharge sources to mimic solar UV radiation, magnetospheric protons, electrons, and galactic cosmic rays \citep[e.g.,][]{Coll1999, Clarke2000, Ramirez2001-la, Szopa2006-iu, Trainer2006-an, Imanaka2007-aj, Sekine+2008, Thissen2009-qf, He2012, Peng2013-oz, Sciamma-OBrien2014-fg, Sebree2014-ob, He2017, Sciamma-OBrien2017-jz, Horst+2018, Sebree2018-yj, Sebree2018-zy, Bourgalais2020-qz, Perrin2021-wz, He2022, Nuevo2022-bg, Hirai2023, Liu2023-vl, Sciamma-OBrien2023-vo, Drant+2026, Husic+2026}. The high-energy irradiation within these experiments leads to the dissociation and ionization of gas species, initiating complex chemical pathways that culminate in the formation of tholin \citep{Cable2012-sf}. Extensive prior work delves into the synthesis environment and evolution of tholins, measuring gas-phase products \citep{Israel2005-nx, Horst+2018, Berry2019-xb, Chatain2023-va}, monitoring how the chemical structure of tholin evolves during synthesis \citep{Sciamma-OBrien2014-fg, Reed2020-rx}, characterizing secondary solid-phase products produced by additional irradiation \citep{Gudipati2013-ak, Cleaves2014-tf}, and studying the degradation of tholin in various environmental conditions \citep{Sciamma-OBrien2017-jz, Chatain2020-rw, Chatain2023-va}. Additionally, significant effort is devoted to measuring the material's bulk properties, including its chemical composition and structure \citep{Derenne2012-jm, Sebree2014-ob, Horst+2018, Sebree2018-zy, Bourgalais2019-hc, Dubois2019-vg, Ruger2019-bm, Dubois2020-dw, Maillard2021-ks, Nuevo2022-bg}, solubility \citep{Carrasco2009-nj, He+2014}, density \citep{Horst2013-ct, Yu_2023-zd}, morphology \citep{Carrasco2009-nj, Sciamma-OBrien2017-jz}, and optical properties \citep{Brucato+2010, Imanaka2012-rm, Gautier2012-ix, Sebree2014-ob, He2022-kb, Sciamma-OBrien2023-vo, Drant+2026}. 

However, it is important to note that specific experimental conditions, including gas composition, pressure, temperature, and energy flux, can significantly influence the physical and chemical properties of laboratory-produced tholins, leading to variations in their particle size distributions, morphology, and chemical composition \citep{Imanaka2004-nt, Sciamma-OBrien2010-qi, Mahjoub+2012, Mahjoub2014-og, Horst+2018, Li2022, Sciamma-OBrien2023-vo}. As a result, there remain open questions regarding how closely laboratory-produced tholins resemble Titan's atmospheric hazes. Only a limited number of studies have systematically examined how tholins produced using different experimental setups and synthesis conditions vary in their properties \citep{coll2013can, cable2014identification, Li2022, bond2025cross, Drant+2026}. 

Building on previous work by \citet{Li2022}, this study examines the surface energy of 32 newly-synthesized tholin samples produced at three laboratories: the Photochemical Aerosol Chamber (PAC) at the University of Northern Iowa (UNI) \citep{Sebree2018-zy}, the Planetary Haze Research Facility (PHAZER) at Johns Hopkins University (JHU) \citep{He2017}, and the Cosmic Simulation Chamber (COSmIC) at NASA Ames Research Center (ARC) \citep{Sciamma-OBrien2017-jz}. Surface energy is a fundamental property governing particle cohesion, adhesion, and wetting behavior, and thus plays an important role in cloud formation and interactions with Titan's surface, including the potential formation of surface films on Titan's hydrocarbon lakes \citep{Yu2017-oh, Cordier2019-dl, Yu2020, Li2022, Yu_2023-zd, Yu2024}. 

Surface energy can be partitioned into dispersive and polar components \citep{Fowkes1964-wj}, which reflect different types of intermolecular interactions. The dispersive component arises from van der Waals forces (London dispersion interactions) and is present in all materials \citep{vanOss+1988}. The polar component arises from specific interactions such as hydrogen bonding and permanent dipole-dipole interactions. This partitioning can be determined using the Owens---Wendt---Rabel---Kaelble (OWRK) method from contact angle measurements with test liquids of known surface tension components \citep{Owens1969, Kaelble1970, Rabel1971}. For Titan's hazes, the dispersive component governs interactions with nonpolar liquids, like methane and ethane, while the polar component may influence interactions with more polar species, such as hydrogen cyanide (HCN) and acetonitrile (CH$_3$CN). Understanding this partitioning helps predict how Titan's aerosols will participate in cloud formation and interactions with the moon's hydrocarbon lakes.

Previous investigations have already provided important constraints on the surface energy of tholin. \citet{Yu2017-oh} quantified the cohesion and adhesion of tholin synthesized with PHAZER at JHU by cold plasma discharge using the sessile drop contact angle method and atomic force microscopy, with implications for sediment transport on Titan. \citet{Yu2020} extended this work by analyzing tholin also synthesized with PHAZER using the contact angle method, but adding a sample made with far-ultraviolet (FUV) radiation, and employing an additional independent verification of the surface energy of plasma tholin using the surface force apparatus. Their results indicated that tholins produced by different energy sources within the same experimental configuration exhibit similar surface energies and cohesive properties, with implications for the role of Titan hazes as cloud condensation nuclei and their interactions with Titan's lakes. More recently, \citet{Li2022} compared the surface energies of seven tholin samples produced using three independent experimental setups, demonstrating that while all samples exhibit high surface energies and strong cohesion, their polar components vary substantially depending on the choice of experimental setup. Among the various experimental conditions, the energy source was identified as the main driver of the variations between the samples. However, the samples used in these previous studies came from storage and were exposed to ambient conditions at some point before measurement, which may have altered their surface properties. 

In the study presented here, all tholin samples analyzed were synthesized specifically for this work, and pristine samples were measured within a week of synthesis to ensure minimally altered surfaces. We further expanded the parameter space by systematically investigating the effects of substrate choice, air exposure, experimental setup, and initial gas composition on the resulting surface energy of tholins. Previous surface energy studies had largely focused on a single initial gas mixture (5\% CH$_4$ in N$_2$) \citep{Yu2017-oh, Yu2020, Li2022}, whereas Titan's atmospheric methane abundance varies significantly with altitude, ranging from ${\sim}1$\% in the upper atmosphere to ${>}5$\% near Titan's surface \citep{Niemann2005, Horst2017, Nixon2024-ji}. Here, we extended our comparative study to N$_2$/CH$_4$ mixtures with methane concentration spanning 0.1--10\%, which encompasses the full range of methane concentrations relevant to Titan's haze-forming part of the atmosphere, as well as previous experimental work.

We describe the sample production, handling, surface energy measurement techniques, and data reduction in Section~\ref{sec:methods}. In Section~\ref{sec:results}, we present the surface energy results and evaluate the effect of substrate choice (Section~\ref{sec:substrate}), air exposure (Section~\ref{sec:exposure}), initial gas mixture (Section~\ref{sec:gasmix}), and experimental setup (Section~\ref{sec:exp_setup}) on the surface energy of the tholin samples. We then discuss the common traits of the tholin samples in Section~\ref{sec:common} and discuss the implications of these experimental results for cloud formation in Titan's atmosphere and haze--lake interactions on its surface. Lastly, in Section~\ref{sec:clouds} we assess which laboratory setups produce samples that most closely resemble Titan's hazes based on observations to-date.

\section{Methods}\label{sec:methods}

\begin{figure*}[!htb]
    \centering
    \captionsetup[subfigure]{
        position=bottom,
        singlelinecheck=true, 
        justification=centering,
        skip=5pt
    }
    
    \begin{subfigure}[t]{0.32\textwidth}
        \centering
        \vspace{0pt}
        \includegraphics[width=\linewidth]{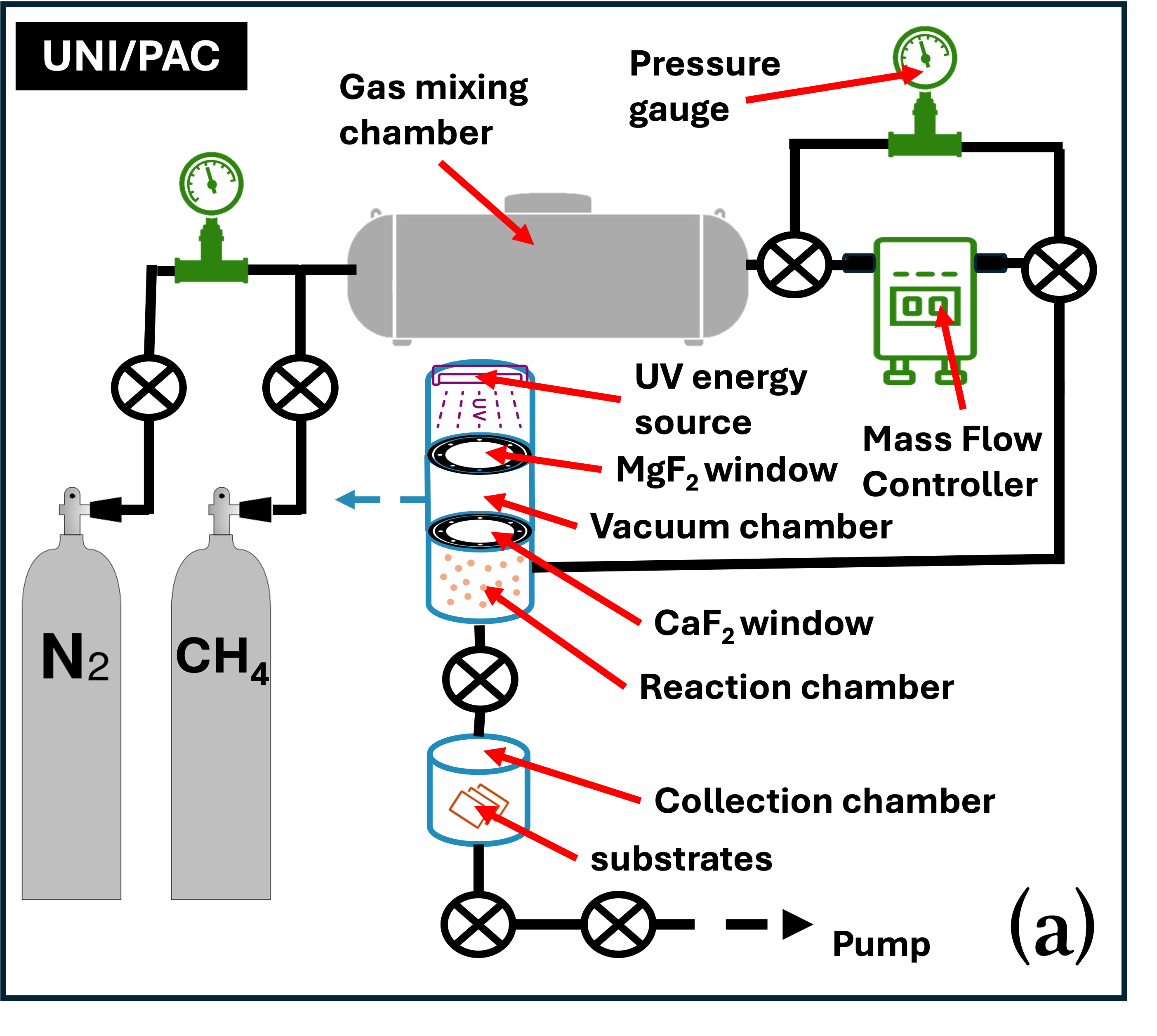}
    \end{subfigure}
    \hfill
    \begin{subfigure}[t]{0.32\textwidth}
        \centering
        \vspace{0pt}
        \includegraphics[width=\linewidth]{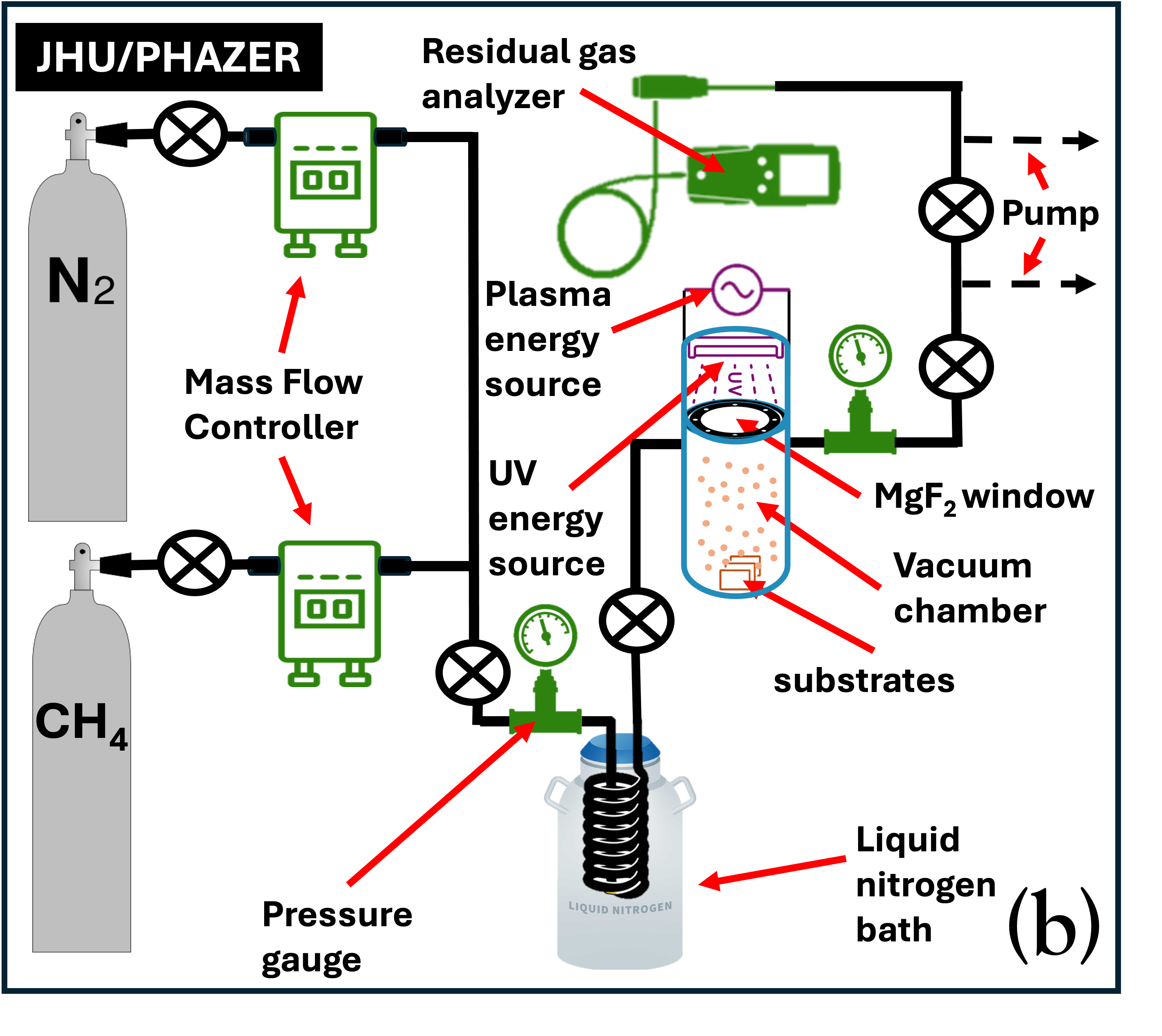}
    \end{subfigure}
    \hfill
    \begin{subfigure}[t]{0.32\textwidth}
        \centering
        \vspace{0pt}
        \includegraphics[width=\linewidth]{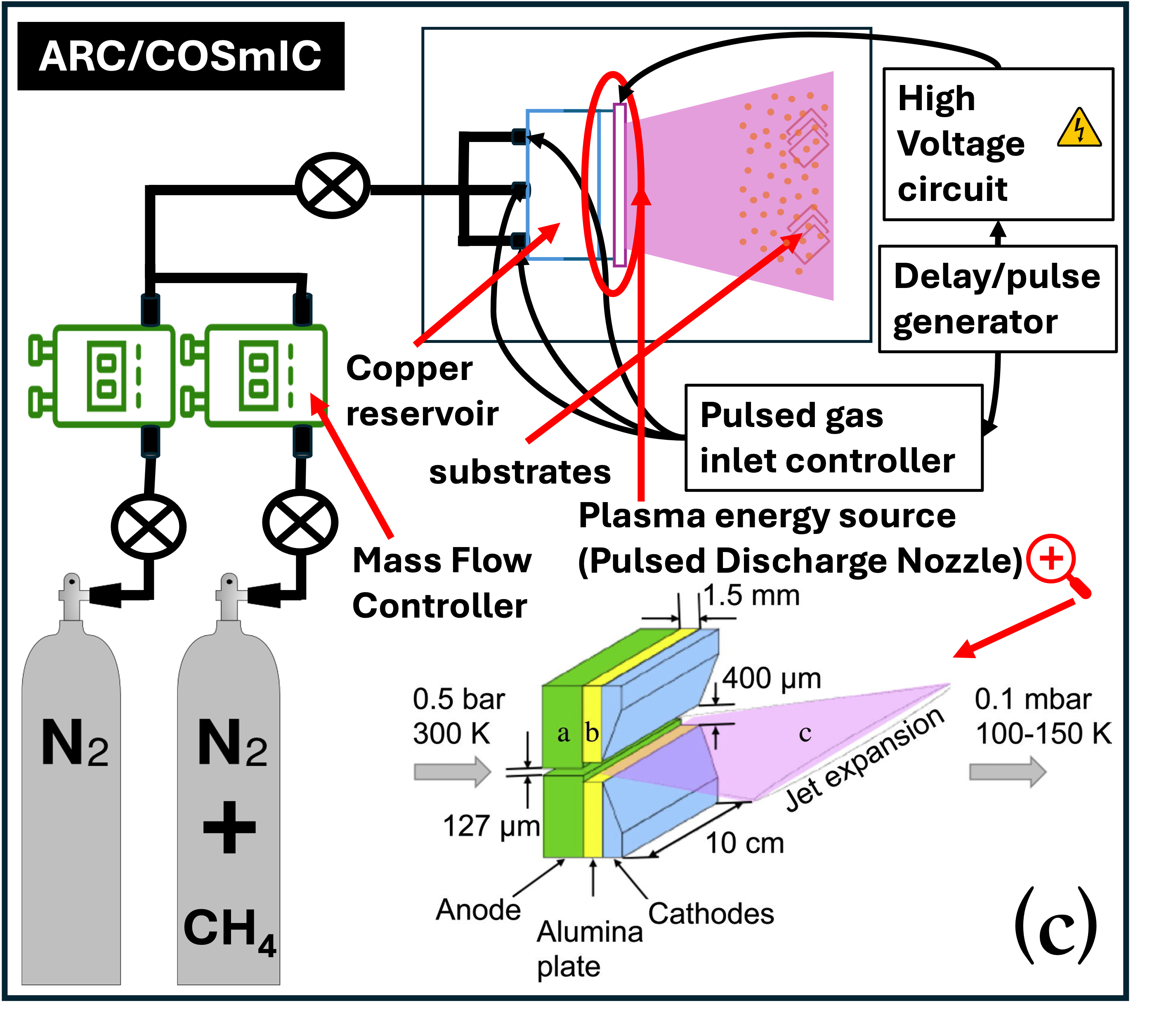}
    \end{subfigure}
    
    \caption{Conceptual diagrams of the three experimental setups used to produce the tholin samples measured in this study: (a) UNI/PAC, (b) JHU/PHAZER, (c) ARC/COSmIC featuring a diagram of the pulsed discharge nozzle adapted from \citet{Sciamma-OBrien2014-fg}.}
    \label{fig1}
\end{figure*}

This study characterized 32 tholin samples produced at three laboratories, each employing distinct experimental setups and conditions (see Table~\ref{table1} and Figure~\ref{fig1}): the PAC chamber at UNI (UNI/PAC), the PHAZER chamber at JHU (JHU/PHAZER), and the COSmIC chamber at NASA ARC (ARC/COSmIC). By comparing tholin samples produced in these laboratories, we directly assess tholins produced via plasma chemistry in two independent laboratories (JHU/PHAZER and ARC/COSmIC), tholin samples produced using FUV photon-driven chemistry in two laboratories (JHU/PHAZER and UNI/PAC), and samples produced using both energy sources in the same setup (JHU/PHAZER plasma and UV). Details of each laboratory configuration and the produced samples are summarized in Section~\ref{sec:production}.

We designed the sample production in two stages. In the first stage, we generated 22 samples using a gas mixture of 5\% CH$_4$ in N$_2$ to systematically investigate the effects of substrate choice and air exposure on the surface energy of the tholin samples. For the substrate comparison, tholins were deposited on three commonly used laboratory substrates: glass slides (Fisherbrand Premium Plain Glass Microscope Slides), quartz discs (Ted Pella, PELCO brand), and mica discs (Ted Pella grade V1 mica). To examine the effect of air exposure, each laboratory produced paired samples on identical substrates, with one sample collected, transferred, and measured with minimal air exposure (hereafter referred to as the ``pristine" sample), and the other sample exposed to ambient laboratory air for approximately 45 days before being measured in the ambient air (``exposed" sample). Details of pristine and exposed sample handling procedures are described in Section~\ref{sec:transport}. 

In the second stage of sample production, we varied the initial gas mixtures to simulate compositional variations in Titan's atmosphere at different altitudes. This led to the production of an additional 10 samples using N$_2$/CH$_4$ mixtures spanning 0.1--10\% CH$_4$, which allowed us to study the dependence of surface energy of tholins as a function of CH$_4$ concentration in the initial gas mixture. For the laboratory setups using plasma discharge (JHU/PHAZER and ARC/COSmIC), we produced tholin samples across four N$_2$/CH$_4$ gas mixtures, containing 1\%, 2\%, 5\%, and 10\% CH$_4$ in N$_2$. For the laboratory setups that use UV lamps (JHU/PHAZER and UNI/PAC), we produced tholin samples with the standard 5\% CH$_4$ in N$_2$ gas mixture as well as an additional lower-methane concentration mixture: a 1\% gas mixture (CH$_4$ in N$_2$) for the JHU/PHAZER UV setup, and a 0.1\% mixture for the UNI/PAC setup. This choice is due to the strong absorption of UV photons by CH$_4$ over the wavelength range of both UV lamps employed in the JHU/PHAZER and UNI/PAC setups, and higher CH$_4$ concentrations leading to rapid attenuation of UV photons and reduced optical penetration depths  \citep[discussed further in Section~\ref{sec:gasmix}, and see also][]{Horst2013-ct}. Because the operating pressure of the JHU/PHAZER chamber is substantially lower than the UNI/PAC chamber (2~Torr versus 500~Torr), we selected a higher CH$_4$ concentration (1\%) for the JHU/PHAZER UV experiment compared to the UNI/PAC UV experiment (0.1\% CH$_4$) in order to achieve comparable methane gas number densities. Even though Titan's present-day atmosphere contains higher CH$_4$ concentrations than these mixtures, the measured sub-nanometer thicknesses of the 5\% CH$_4$ in N$_2$ UV tholin samples (see Section~\ref{sec:ellipsometry}) motivated the use of lower methane concentrations to optimize production rates with the attempt to create thicker samples, thereby reducing thickness-dependent effects in the surface energy measurements.

A complete list of the 32 samples and their production condition is provided in Table~\ref{table1}. We identify the samples by their corresponding table numbers throughout the paper.

\begin{table*}[!htb]
\centering
\caption{List of the 32 tholin samples produced for this study, including details of the experimental setups in which they were created and the associated production and analysis conditions.}
\resizebox{\textwidth}{!}{
\begin{tabular}{|p{7.6cm}cccc|} % Remove adjustbox and use a standard tabular environment
\hline
\multicolumn{1}{|c}{Laboratory/Experimental Setup} & Sample \# & Pristine/Exposed & Substrate & Gas Mix Ratio (N$_2$/CH$_4$) \\
\hline
    JHU/PHAZER & 1     & Pristine & glass slide & 95/5 \\
    \quad Energy Source: AC Plasma Discharge \newline\hspace*{1em}(power source 6~kV, 10~mA) & 2     & Pristine & quartz disk & 95/5 \\
    \quad Reaction Time: 72~hr & 3     & Pristine & mica  & 95/5 \\
    \quad Gas Exposure Time: 3~s & 4     & Exposed & glass slide & 95/5 \\
    \quad Gas Temperature: 100~K & 5     & Exposed & quartz disk & 95/5 \\
    \quad Gas Flow Rate: 10~sccm & 6     & Exposed & mica  & 95/5 \\
    \quad Pressure: 2~Torr (2.67~mbar) & 7     & Pristine & glass slide & 90/10 \\
          & 8     & Pristine & glass slide & 98/2 \\
          & 9     & Pristine & glass slide & 99/1 \\
    \hline
    JHU/PHAZER & 10    & Pristine & glass slide & 95/5 \\
    \quad Energy Source: Hydrogen UV lamp ($115-400$~nm) & 11    & Pristine & quartz disk & 95/5 \\
    \quad Reaction Time: 96~hr (95/5), 144~hr (99/1) & 12 & Pristine & mica  & 95/5 \\
    \quad Gas Exposure Time: 3~s & 13    & Exposed & glass slide & 95/5 \\
    \quad Gas Temperature: 100~K & 14    & Exposed & quartz disk & 95/5 \\
    \quad Gas Flow Rate: 10~sccm & 15    & Exposed & mica  & 95/5 \\
    \quad Pressure: 2~Torr (2.67~mbar) & 16    & Pristine & glass slide & 99/1 \\
          & 17    & Pristine & quartz disk & 99/1 \\
    \hline
    ARC/COSmIC & 18    & Pristine & glass slide & 95/5 \\
    \quad Energy Source: DC Pulsed Plasma Discharge \newline\hspace*{1em}(-700~V, 600~$\mu$s pulse, avg. electron energy 2-4~eV) & 19    & Pristine & quartz disk & 95/5 \\
    \quad Reaction Time: 9.75-11.75~hr & 20    & Pristine & mica  & 95/5 \\
    \quad Gas Exposure Time: 3.5~$\mu$s & 21    & Exposed & glass slide & 95/5 \\
    \quad Gas Temperature: 150~K & 22    & Exposed & quartz disk & 95/5 \\
    \quad Gas Flow Rate: 2000~sccm & 23    & Exposed & mica  & 95/5 \\
    \quad Pressure: 22.5~Torr (30.0~mbar) & 24    & Pristine & glass slide & 90/10 \\
          & 25    & Pristine & glass slide & 98/2 \\
          & 26    & Pristine & glass slide & 99/1 \\
          & 27    & Pristine & quartz disk & 99/1 \\
    \hline
    UNI/PAC & 28    & Pristine & quartz disk & 95/5 \\
    \quad Energy Source: Deuterium lamp ($115-400$~nm) & 29    & Pristine & mica  & 95/5 \\
    \quad Reaction Time: 162.5~hr (95/5), 421~hr (99.9/0.1) & 30    & Exposed & quartz disk & 95/5 \\
    \quad Gas Exposure Time: 180~s & 31    & Exposed & mica  & 95/5 \\
    \quad Gas Temperature: 300~K & 32    & Pristine & glass slide & 99.9/0.1 \\
    \quad Gas Flow Rate: 20~sccm &       &       &       &  \\
    \quad Pressure: 500~Torr (667~mbar) &       &       &       &  \\
    \hline
\end{tabular}
}
\label{table1}
\end{table*}

\subsection{Production of Tholin Samples} \label{sec:production}
In all three experimental setups, tholin synthesis proceeds via the formation of solid aerosol particles in the gas phase, which are then deposited onto various substrates for analysis. However, the setups differ in their collection geometry: in the JHU/PHAZER facility, particles are deposited onto substrates located within the main reaction chamber, whereas in the UNI/PAC and ARC/COSmIC facilities, the particles are transported by the gas flow to be deposited on substrates located downstream of the primary reaction zone.

Here, we briefly describe the tholin sample production at each laboratory. Detailed descriptions of the UNI/PAC, JHU/PHAZER, and ARC/COSmIC facilities can be found in \citet{Sebree2018-zy}, \citet{He2017}, and \citet{Sciamma-OBrien2014-fg}, respectively.

At UNI, tholin samples were produced within the PAC chamber \citep{Sebree2014-ob, Sebree2018-zy, Sebree2018-yj}, employing a Lyman-$\alpha$ UV deuterium lamp (Hamamatsu L11798) as the energy source. The lamp emitted radiation in the $115-400$~nm wavelength range with a photon flux of approximately 5.0$\times$10$^{15}$~photons~s$^{-1}$. N$_2$ and CH$_4$ gases were delivered separately via mass flow controllers to a 2~L stainless-steel gas mixing manifold where they sat for a minimum of 8 hours until homogeneity was reached. The homogenized gas mixture then continuously flowed into a UV-reaction chamber with a MgF$_2$ window, where photochemistry ensued for $\sim$162.5~hr for the 5\% CH$_4$ in N$_2$ gas mixture and $\sim$421~hr for the 0.1\% gas mixture. The resulting aerosols, formed at 300~K and 500~Torr, were carried by the flow of gas from the reaction chamber downstream and deposited onto the collection substrates in a separate collection chamber. 

UNI provided a total of 5 samples for this study. For the 5\% CH$_4$ in N$_2$ gas mixture, two experimental runs were performed (one pristine, one exposed), with substrates deposited simultaneously in each run. Due to the limited size of the UNI/PAC collection chamber, only two substrates, one mica disc and one quartz disc, could be accommodated simultaneously; a glass slide was therefore excluded from these runs. Of the four samples produced, one mica and one quartz sample were kept pristine, while the remaining two were exposed to air. Due to the extremely long experiment time required for the lower methane concentration ($\sim$421~hr), only a single run was feasible for the 0.1\% CH$_4$ in N$_2$ gas mixture. For this run, a single glass slide was placed in the collection chamber. This sample was kept pristine and used for the gas mixture comparisons.

At JHU, the PHAZER chamber \citep{He2017} synthesized tholin samples using two energy sources: a hydrogen UV lamp with an MgF$_2$ window (HHeLM-L, Resonance Ltd.) and AC glow plasma discharge. The UV lamp emits wavelengths between $115-400$~nm with a photon flux of approximately 1.4$\times$10$^{15}$~photons~s$^{-1}$. N$_2$ and CH$_4$ gases were delivered separately via mass flow controllers at a flow rate of 10~standard cubic centimeters per minute (sccm). The homogeneous gas mixture was cooled to approximately 100~K by passing through a 15~m stainless-steel coil immersed in a liquid nitrogen cold bath (77~K), before being exposed to one of the two energy sources at a chamber pressure of 2~Torr. Tholins were produced within the reaction chamber for 96-144~hr for the UV samples and $\sim$72~hr for the plasma samples and deposited onto the substrates within the chamber.

For this study, JHU/PHAZER produced a total of 17 samples. For the plasma energy source, six samples were produced using the 5\% CH$_4$ in N$_2$ gas mixture (two each on mica, quartz, and glass substrates), with half of the samples kept pristine and half exposed to air. These six samples were produced in two simultaneous deposition runs, one pristine and one exposed, with all three substrate types deposited together in each run. Three additional pristine plasma samples were produced using CH$_4$ concentrations of 1\%, 2\%, and 10\% in N$_2$ on glass slide substrates. For the UV energy source, six samples were produced on the same three substrate types, again with half being pristine and half being exposed. Two additional pristine UV samples were produced using the 1\% CH$_4$ in N$_2$ gas mixture on glass and quartz substrates.

At NASA ARC, tholin samples were synthesized using the COSmIC chamber \citep{Sciamma-OBrien2017-jz}, which uses DC plasma discharge as the energy source. A pulsed discharge nozzle \citep{2006CP....326..445B} injected premixed gas at a flow rate of 2000~sccm into a vacuum chamber. A supersonic jet expansion, resulting from the injection of the gas through a thin slit, cooled the gas to approximately 150~K and reduced the pressure to 22.5~Torr. A 700~V potential applied to cathodes flanking the slit generated the DC plasma discharge within the gas stream, inducing chemistry. Experimental run times ranged from 9--12~hr. The produced solid aerosol particles were transported by the gas jet and deposited onto substrates positioned approximately 5~cm downstream from the electrodes.

ARC/COSmIC provided a total of 10 samples for this study. Six samples were produced using the 5\% CH$_4$ in N$_2$ gas mixture on glass, quartz, and mica substrates, with half of the samples kept pristine and half exposed to air. These six samples were produced in two simultaneous deposition runs, one pristine and one exposed, with all three substrate types deposited together in each run. Four additional pristine samples were produced using CH$_4$ concentrations of 1\%, 2\%, and 10\% in N$_2$ on glass slide substrates, along with one additional pristine 1\% sample deposited on a quartz disc.

\subsection{Sample Collection, Transportation, Storage, and Measurement Procedures} \label{sec:transport}

\begin{figure}[!htb]
\centering
\includegraphics[width=\columnwidth]{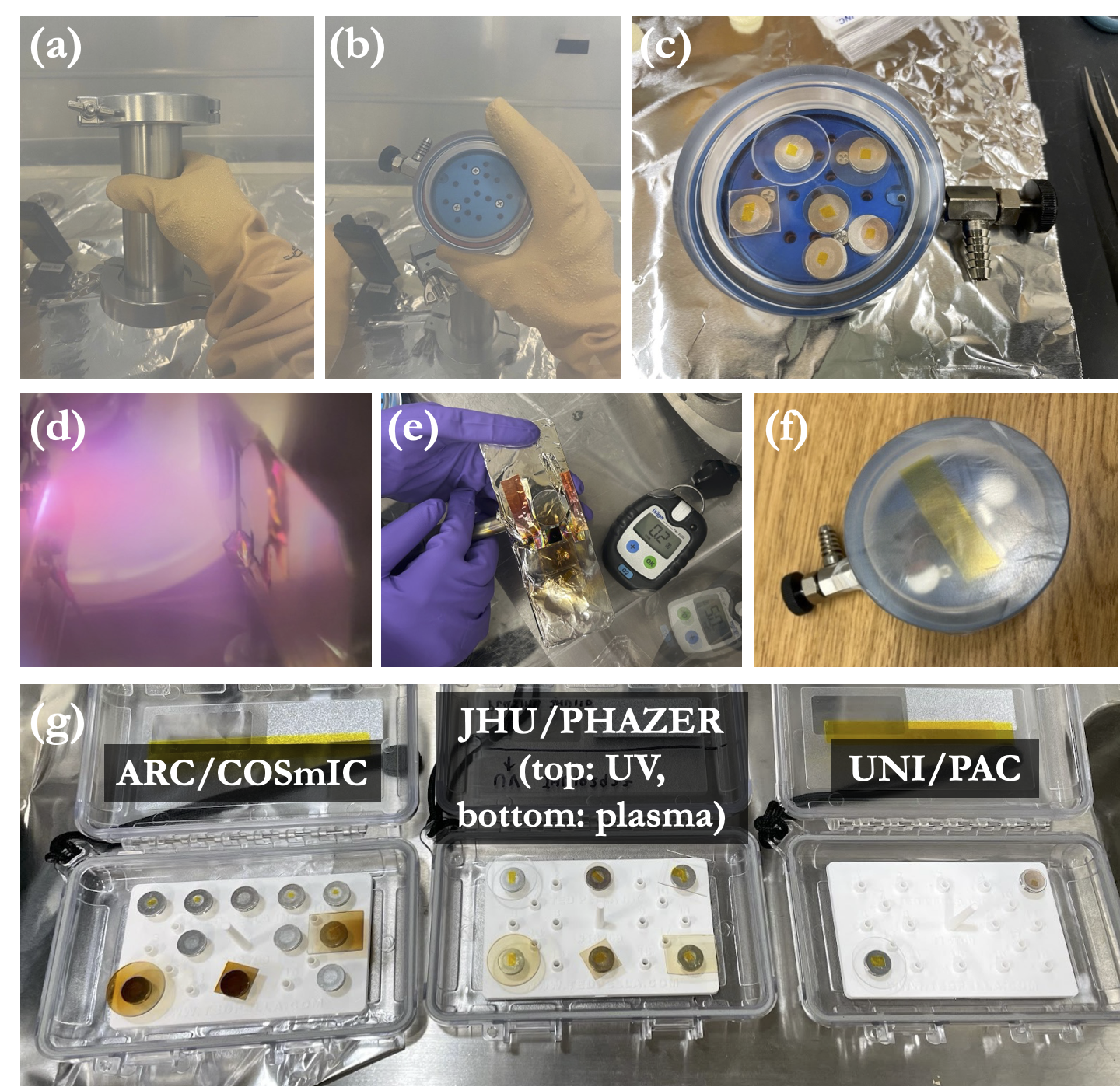}
\caption{Images showing the pristine sample production and handling process with the ARC/COSmIC setup samples as an example (panels a-f) and all the exposed samples (panel g). (a) Custom-built dry N$_2$ transporting vessel handled in the N$_2$ glove box. (b) PELCO vacuum desiccator used for sample transportation handled in the N$_2$ glove box. (c) PELCO vacuum desiccator containing blank substrates prior to sample deposition. (d) Tholin samples being produced in the ARC/COSmIC setup. (e) Tholin sample being handled and packed in an N$_2$ glove box with an oxygen sensor showing 0.2\% vol O$_2$ level. (f) Packed PELCO vacuum desiccator containing pristine tholin samples under vacuum and ready for shipment. (g) Exposed samples shipped in PELCO SEM pin storage boxes (from left to right: ARC/COSmIC, JHU/PHAZER with UV samples on the top and plasma samples on the bottom, and PAC/UNI). All samples shown were produced using the 5\% CH$_4$ in N$_2$ gas mixture.}
\label{fig:lab}
\end{figure}

To ensure a fully comparable study across all laboratories, tholin samples were collected, transported, stored, and measured following a standardized method developed in this study. Quartz and glass substrates were prepared at the Planetary Material Characterization Facility (PMCHEF) at UT San Antonio prior to shipment to the tholin production laboratories. These substrates were cleaned using the following procedure to ensure dust- and organic-free surfaces for sample deposition: (1) soak in piranha solution (3:1 sulfuric acid to hydrogen peroxide by volume) for 2~hr to remove organic contaminants; (2) Rinse and sonicate sequentially in deionized water to remove piranha solution, then in acetone and isopropyl alcohol (IPA) for 15~min each to remove residual weakly-bonded organics and moisture; and (3) soak in IPA before being taken out for shipment to each tholin production laboratory. Cleaned substrates were stored in a sealed beaker of IPA until they were ready for shipment. When ready, cleaned substrates were held in a PELCO vacuum-desiccator under vacuum for $\sim$2 hours before being sealed in the desiccator (Figure~\ref{fig:lab}c). The desiccator was then wrapped in parafilm and secured in a shipping box. Mica substrates were not subjected to the piranha cleaning procedure; instead, they were prepared by mechanical cleavage at each tholin production laboratory immediately prior to deposition, creating a molecularly smooth surface with some step edges over large areas, following the procedure described in \citet{Yu2020}. 

After tholin deposition (an example of the sample deposition process in action is shown for the ARC/COSmIC setup in Figure~\ref{fig:lab}d), the samples were categorized as either ``pristine" or ``exposed". Pristine samples were collected and packed entirely within N$_2$ glove boxes at each respective laboratory, ensuring they were never exposed to ambient air (an example of sample handling in the N$_2$ glove box is shown for the ARC/COSmIC setup in Figure~\ref{fig:lab}e, and the packed samples under vacuum are shown in Figure~\ref{fig:lab}f). They were then shipped overnight to PMCHEF in custom-built transport vessels under inert conditions, either a dry nitrogen transport container built with a Kurt~J.~Lesker KF40 flange or a vacuum desiccator (PELCO), see Figure~\ref{fig:lab}(a)-(b). Upon receipt, pristine samples were stored in a vacuum desiccator and handled in a dry nitrogen environment maintained within a glove box, see Figure~\ref{fig:lab}(a)-(b). All measurements of the pristine samples were performed within one week of synthesis. The measurements were performed under controlled conditions inside a glove box purged with high-purity dry nitrogen (99.999\%), achieving a relative humidity (RH) $<$0.1\%. This low-RH environment was maintained for at least 30~minutes before the start of the measurement and continuously monitored throughout the experiment.

Exposed samples were shipped in PELCO SEM pin storage boxes (Figure~\ref{fig:lab}g). Upon arrival, these samples remained in their storage boxes in ambient laboratory air ($\sim$40\% RH) for ${\sim}45$ days prior to measurement. All measurements of the exposed samples were conducted under ambient laboratory conditions at room temperature, with RH ranging between 14\% and 39\%.

\subsection{Surface Energy Determination}
\label{sec:SE}
The surface energy of tholin samples can be determined using contact angle measurements with two probe liquids of known surface tensions. A detailed description of the theoretical framework is provided in \citet{Yu2020} and \citet{Li2022}. Here, we briefly summarize the approach adopted in this study. 

We employ the OWRK method, also known as the geometric mean method, to determine the surface energy of the solid samples. This method assumes that the total surface energy of a solid, or the surface tension of a liquid ($\gamma^{tot}$), is composed of a dispersive component ($\gamma^{d}$) and a polar component ($\gamma^{p}$), such that $\gamma^{tot} = \gamma^{d} + \gamma^{p}$. The dispersive component accounts for London dispersive forces, while the polar component represents dipole-dipole and hydrogen-bonding interactions \citep{Owens1969, Kaelble1970, Rabel1971}. For a given liquid-solid pair, the OWRK method gives:
\begin{equation}
\label{eqn3}
\gamma_{lv}^{tot}(1+\cos\theta)=2(\sqrt{\gamma^{d}_{sv}\gamma^{d}_{lv}}+\sqrt{\gamma^{p}_{sv}\gamma^{p}_{lv}}),
\end{equation}
where $\gamma_{lv}^{tot}$, $\gamma^{d}_{lv}$, and $\gamma^{p}_{lv}$ are the total, dispersive, and polar components of the liquid surface tension, respectively, $\gamma^{d}_{sv}$ and $\gamma^{p}_{sv}$ are the dispersive and polar components of the solid surface energy, and $\theta$ is the measured contact angle between the liquid and the solid surface.

Because Equation~\ref{eqn3} contains two unknowns ($\gamma^{d}_{sv}$ and $\gamma^{p}_{sv}$), contact angle measurements with two probe liquids, one non-polar and one polar, are required, turning Equation~\ref{eqn3} into a set of two equations (one for each test liquid) that can be solved for the dispersive and polar components of the solid surface energy. The total surface energy is then obtained as $\gamma^{tot}_{sv} = \gamma^{d}_{sv} + \gamma^{p}_{sv}$. In this study, the weighted mean contact angles and their associated uncertainties (Section~\ref{contactAngleMeasurements}) were used as inputs to Equation~\ref{eqn3}. The full analytical solutions of the OWRK method for $\gamma^{tot}_{sv}$, $\gamma^{d}_{sv}$, and $\gamma^{p}_{sv}$ and their 1$\sigma$ uncertainties can be found in Appendix~A of \citet{Li2022}.

\subsection{Contact Angle Measurements}
\label{contactAngleMeasurements}
% As described in Section~\ref{sec:SE}, we need the contact angles between the solid material and two test liquids with known surface tensions to determine the surface energy of the tholin samples. 
Contact angle measurements were performed using an Ossila goniometer and its associated analysis software, which enables image and video capture as well as automated contact angle fitting. We used the sessile drop method, where the tholin-coated substrate was placed horizontally, and a liquid droplet was dispensed onto the surface using a microliter syringe \citep{Drelich2013}. Each measurement was recorded as a 60~s video at 30~frames~s$^{-1}$ for subsequent analysis. A drop volume of less than 2~$\mu$L was used for all measurements. This small volume minimized the influence of gravitational distortion on the drop shape, providing a more accurate representation of the intrinsic contact angle \citep{Zhang2008-ki, Extrand2010-kv, Yu2020} and allowed multiple replicate measurements on the limited surface area of each sample. The Ossila software applies two fitting algorithms to determine the contact angle: circle fitting for angles less than 10$^\circ$, which assumes a circular droplet profile, and polynomial fitting for larger contact angles, which provides a more accurate representation of the droplet shape.

Following \citet{Hejda2010} and \citet{Yu2020}, we use two established test liquids, diiodomethane ($>$99\%; ARCOS Organics) and high-performance liquid chromatography (HPLC) grade water (Fisher Chemical), to measure the surface energy of the tholin samples. The surface tensions and the corresponding dispersive and polar components of these liquids are summarized in \citet[][Table~2]{Li2022}. For each sample, we measured the contact angles at a minimum of five distinct locations using each test liquid. We measured contact angles with diiodomethane first for each sample before switching to water to minimize the effects of water adsorption on the tholin samples. Before switching to water, any remaining diiodomethane droplets were carefully absorbed using a Kimwipe. Furthermore, the water droplets were always dispensed on well-separated surface areas, away from the locations where the diiodomethane droplets had been deposited, to prevent any cross-contamination between the two test liquids.

\begin{figure*}[!htb]
    \centering
    \captionsetup[subfigure]{
    position=bottom,
    singlelinecheck=true, 
    justification=centering,
    skip=5pt
}
    \begin{subfigure}{0.46\textwidth}
        \includegraphics[width=\textwidth]{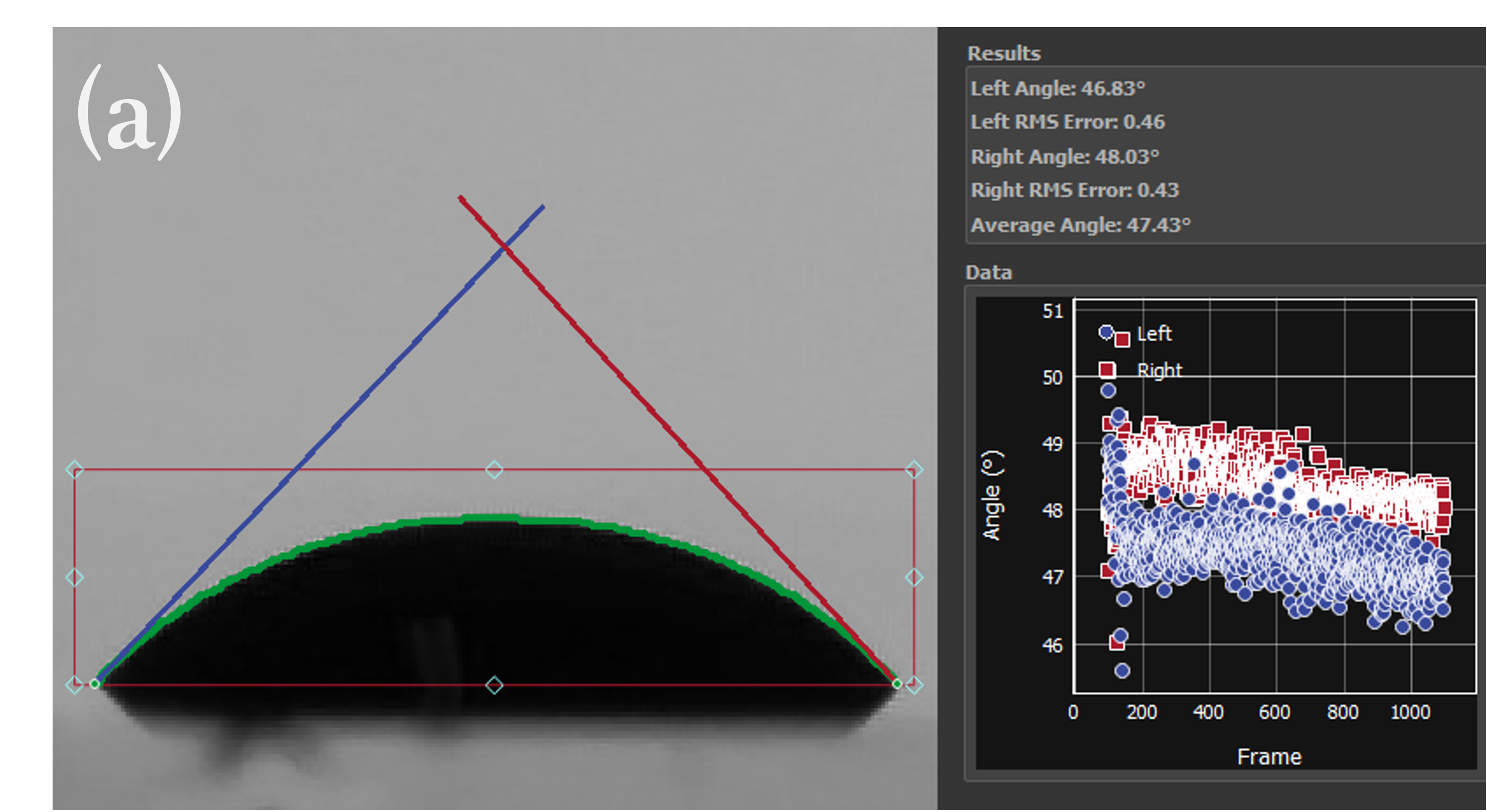}
        
    \end{subfigure}
    \hfill
    \begin{subfigure}{0.49\textwidth}
        \includegraphics[width=\textwidth]{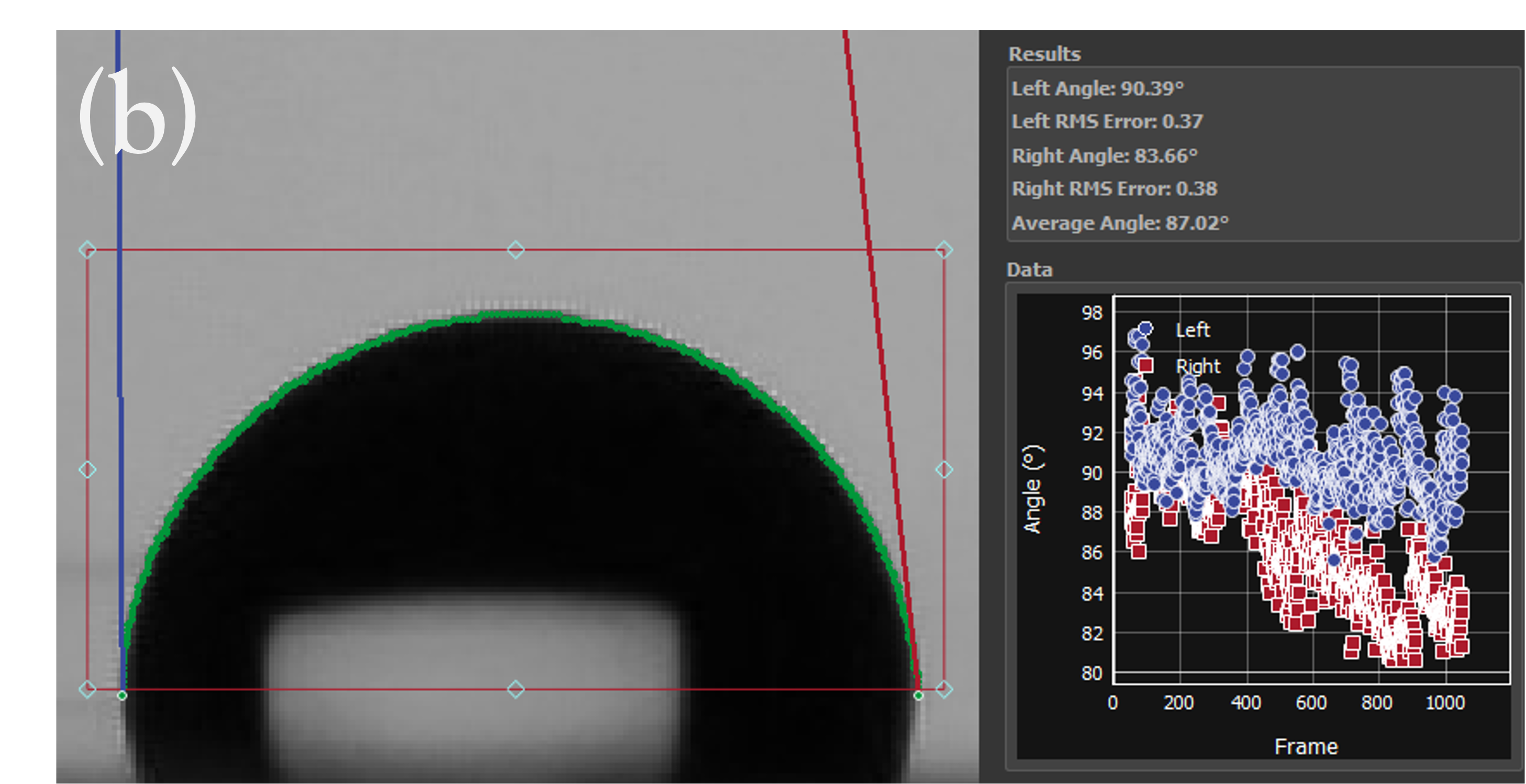}
        
    \end{subfigure}
    \hfill
    \caption{Representative contact angle measurement interface from the Ossila goniometer software for tholin produced at ARC/COSmIC with 10\% CH$_4$ in N$_2$ on glass. Each panel shows a still frame (left) with the region of interest box (red) and fitted contact angles (blue line: left angle; red line: right angle), alongside numerical data (top right) displaying the left angle, right angle, and their average with root mean square errors, and a time series plot (bottom right) of contact angle versus frame number. (a) Diiodomethane droplet showing consistent contact angle measurements over time (angles range 45$^{\circ}$--51$^{\circ}$) with a smooth trend in the time series. (b) Water droplet showing greater variability in contact angle measurements over time (angles range 80$^{\circ}$--98$^{\circ}$) with more scatter in the time series, likely due to faster evaporation and possible dissolution effects \citep{Li2022}.}
    \label{fig2}
    
\end{figure*}
\begin{table*}[!htb]
  \centering
  \caption{The measured contact angles and 1$\sigma$ uncertainties between tholin samples and test liquids. Contact angles for the bare glass, quartz, and mica substrates are included as reference values to facilitate assessment of substrate influence on the tholin surface energy measurements. $^a$Water droplets on blank mica wetted completely, and no contact angle could be directly measured; a value of $5\pm5^\circ$ was adopted, consistent with the software's classification of low angles below $10^\circ$.}
  \scriptsize
    \begin{tabularx}{\textwidth}{l|Y|Y}
    \toprule
    \multicolumn{1}{l|}{Sample} & \multicolumn{2}{c}{\begin{tabular}[c]{c}Test Liquid Contact Angle ($^\circ$)\end{tabular}} \\
    \midrule
          & Diiodomethane & HPLC Water \\
    \midrule
    1 (95/5 Plasma)      & 28.92~$\pm$~1.01 & 26.13~$\pm$~2.28 \\
    2 (95/5 Plasma)      & 29.92~$\pm$~0.85 & 18.04~$\pm$~1.86 \\
    3 (95/5 Plasma)      & 32.56~$\pm$~0.91 & 23.47~$\pm$~4.65 \\
    4 (95/5 Plasma)      & 43.63~$\pm$~0.59 & 29.10~$\pm$~3.30 \\
    5 (95/5 Plasma)      & 47.62~$\pm$~1.15 & 27.37~$\pm$~5.67 \\
    6 (95/5 Plasma)      & 45.71~$\pm$~0.96 & 28.06~$\pm$~5.52 \\
    7 (90/10 Plasma)     & 32.90~$\pm$~0.62 & 27.95~$\pm$~1.20 \\
    8 (98/2 Plasma)      & 45.96~$\pm$~0.65 & 23.74~$\pm$~1.76 \\
    9 (99/1 Plasma)      & 38.13~$\pm$~1.82 & 29.52~$\pm$~1.30 \\
    \midrule
    10 (95/5 UV)         & 51.49~$\pm$~1.21 & 67.54~$\pm$~2.14 \\
    11 (95/5 UV)         & 51.45~$\pm$~0.85 & 61.14~$\pm$~2.24 \\
    12 (95/5 UV)         & 62.41~$\pm$~1.34 & 75.79~$\pm$~1.77 \\
    13 (95/5 UV)         & 58.62~$\pm$~0.38 & 81.60~$\pm$~1.19 \\
    14 (95/5 UV)         & 55.15~$\pm$~1.80 & 73.47~$\pm$~2.00 \\
    15 (95/5 UV)         & 59.76~$\pm$~0.92 & 61.53~$\pm$~1.09 \\
    16 (99/1 UV)         & 45.68~$\pm$~0.83 & 54.48~$\pm$~1.80 \\
    17 (99/1 UV)         & 52.42~$\pm$~0.73 & 59.10~$\pm$~0.36 \\
    \midrule
    18 (95/5 Plasma)     & 23.09~$\pm$~1.50 & 73.02~$\pm$~3.16 \\
    19 (95/5 Plasma)     & 19.65~$\pm$~2.53 & 72.64~$\pm$~2.12 \\
    20 (95/5 Plasma)     & 14.51~$\pm$~1.33 & 78.07~$\pm$~5.79 \\
    21 (95/5 Plasma)     & 46.80~$\pm$~0.89 & 66.70~$\pm$~2.14 \\
    22 (95/5 Plasma)     & 46.66~$\pm$~1.51 & 66.79~$\pm$~2.22 \\
    23 (95/5 Plasma)     & 47.08~$\pm$~0.89 & 63.08~$\pm$~1.91 \\
    24 (90/10 Plasma)    & 49.60~$\pm$~1.69 & 86.50~$\pm$~2.43 \\
    25 (98/2 Plasma)     & 50.90~$\pm$~1.10 & 88.63~$\pm$~1.09 \\
    26 (99/1 Plasma)     & 40.83~$\pm$~0.49 & 84.67~$\pm$~1.55 \\
    27 (99/1 Plasma)     & 39.12~$\pm$~1.66 & 76.51~$\pm$~1.60 \\
    \midrule
    28 (95/5 UV)         & 54.81~$\pm$~0.95 & 32.19~$\pm$~1.15 \\
    29 (95/5 UV)         & 49.28~$\pm$~0.38 & 27.49~$\pm$~2.69 \\
    30 (95/5 UV)         & 54.55~$\pm$~0.63 & 41.26~$\pm$~2.79 \\
    31 (95/5 UV)         & 50.69~$\pm$~1.34 & 25.41~$\pm$~0.96 \\
    32 (99.9/0.1 UV)     & 55.00~$\pm$~0.72 & 41.17~$\pm$~3.72 \\
    \midrule
    Blank Glass          & 36.57~$\pm$~1.40 & 15.61~$\pm$~2.47 \\
    Blank Quartz         & 39.35~$\pm$~1.22 & 21.73~$\pm$~0.81 \\
    Blank Mica           & 34.06~$\pm$~1.64 & $5.00~\pm~5.00^a$ \\
    \bottomrule
    \end{tabularx}%
  \label{table3}%
\end{table*}%

The Ossila software allows us to extract a fitted contact angle from each video frame, thus generating a frame-resolved contact angle time series for each droplet (Figure~\ref{fig2}). Video recording began just before the deposition of the drop onto the substrate. The analysis time window was selected based on the probe liquid and sample type. For diiodomethane droplets, the full 60-second video was analyzed for all samples due to the minimal evaporation of diiodomethane and the negligible dissolution of tholin samples in diiodomethane within this timeframe. For water droplets, shorter analysis windows were required to minimize the effects of evaporation and tholin film dissolution. Specifically, the first 5~s of the video were analyzed for UV samples, while an even shorter window (first 2.5~s) was used for plasma samples due to their faster dissolution rates \citep{Li2022}. 

We note that using water as a probe liquid introduces a potential complication. Plasma-synthesized tholins are known to contain kinetically trapped free radicals \citep{Budil+2003}, and these reactive surface sites could, in principle, react with the water droplet during measurement, modifying the surface chemistry in real time. The short measurement windows employed here (2.5--5~s) minimize this effect, and laboratory hydrolysis studies indicate that significant chemical modification of tholins in liquid water requires timescales of days to weeks \citep{Neish+2009, Neish+2010}. Nevertheless, because any immediate quenching of highly reactive surface sites by the water droplet would passivate the surface and reduce its apparent polarity \citep{Liston+1993, Kondyurin+2009}, the polar component values reported here may represent a conservative lower bound on the true surface reactivity of freshly synthesized pristine tholins. This effect may differ in magnitude between experimental setups if the density of trapped radicals varies, meaning that inter-laboratory differences in the polar component could be partially influenced by differences in radical quenching rates during measurement. Substituting water with alternative polar probe liquids (e.g., formamide or ethylene glycol) is not a viable solution, as \citet{Yu2020} demonstrated that tholins are also soluble in these liquids, preventing stable contact angle measurements. Future studies could potentially circumvent this limitation by employing inverse gas chromatography, which determines both dispersive and polar surface energy components through transient gas-phase probe molecule adsorption without bulk liquid-surface contact \citep{Klein+2015}, thereby reducing the likelihood of radical quenching during measurement.
 
For each droplet $i$, a representative contact angle $\theta_i$ was obtained by averaging the frame-resolved fitted contact angles over the selected time window, with an associated standard deviation, $\sigma_{\theta_i}$. For each test liquid, we dispensed multiple droplets on a given sample, and we calculated the mean contact angle ($\overline\theta$) using an inverse-variance-weighted average:
\begin{equation}
    \overline\theta = \frac{\sum_{i=1}^{n} w_i\theta_i}{\sum_{i=1}^{n} w_i}, w_i = \frac{1}{\sigma_{\theta_i}^2}.
\end{equation}
The corresponding uncertainty of the weighted mean contact angle was calculated as:
\begin{equation}
    \sigma_\theta = \sqrt{\frac{1}{\sum_{i=1}^{n} w_i}}.
\end{equation}
This expression, however, accounts only for frame-to-frame variation within each droplet and does not include any additional droplet-to-droplet scatter. To incorporate this, we first computed the reduced chi-square:
\begin{equation}
    \chi_\nu^2 = \frac{1}{n-1}\sum_i w_i(\theta_i-\overline\theta)^2,
\end{equation}
where $n$ is the number of droplets per test liquid on a given sample. The final uncertainty, incorporating both within-droplet (frame-to-frame variations) and between-droplet variations, was then:
\begin{equation}
    \sigma_{\theta,\mathrm{final}} = \sigma_\theta\sqrt{\mathrm{max}(1,\chi_\nu^2)}.
\end{equation}

The calculated weighted mean contact angle and corresponding 1$\sigma$ uncertainties for each test liquid on each sample are summarized in Table~\ref{table3}.

To assess whether two samples differ in surface energy, we treated the derived surface energy values as independent estimates and quantified the statistical significance of their difference. The difference of the two surface energy values $\Delta\gamma=\gamma_1-\gamma_2$ and its uncertainty $\sigma_{\Delta\gamma} = \sqrt{\sigma_{\gamma1}^2+\sigma_{\gamma2}^2}$ were used to compute a z-score, 
\begin{equation}
    z=\frac{|\Delta\gamma|}{\sigma_{\Delta\gamma}}.
\end{equation} 
Values of $z\geq3$ (corresponding to a two-sided p-value $p\leq0.003$) were interpreted as statistically distinguishable. In the discussion below (Section~\ref{sec:results}), we comment both on the statistical significance and on the absolute magnitude of the difference in mJ~m$^{-2}$.

\subsection{Spectroscopic Ellipsometry to Determine Thickness of the Samples}\label{sec:ellipsometry}
To assess whether deviations in surface energy behavior observed for a subset of samples could be attributed to unusually low tholin film thicknesses, we performed spectroscopic ellipsometry (SE) measurements on a representative subset of tholin samples. SE measurements were carried out using a J.A. Woollam M2000-DI variable-angle spectroscopic ellipsometer at PMCHEF, and data were analyzed in CompleteEASE software (J.A. Woollam Co.). Measurements were performed at three angles of incidence to improve sensitivity to film thickness and minimize parameter correlation during fitting. SE measurements were performed after the contact angle characterization, and the precise SE laser was targeted exclusively on areas of the film that were visually unaffected by the contact angle test liquids to ensure accuracy.

In SE, changes in the polarization state of reflected light are measured and analyzed through the parameters $\Psi$ (related to the amplitude ratio) and $\Delta$ (related to the phase difference) to extract physical parameters such as film thickness. Because SE measurements are sensitive to both the film and underlying materials, each substrate was characterized independently with SE to determine its optical constants and surface roughness. Following substrate characterization, parameters related to the substrate were fixed, and tholin thickness was determined by fitting ellipsometry data in the transparent spectral region, where absorption is negligible ($600-1700$~nm) and thickness can be reliably decoupled from optical constants. The tholin film was assumed to form a single layer, and its refractive index was described using the Cauchy empirical equation \citep{TompkinsIrene2005}. Due to their extremely low film thicknesses, all UV-generated samples were modeled as a simple roughness layer, as fitting for both optical constants and thickness simultaneously was not feasible. The results of these measurements for a subset of exposed samples on quartz are presented in Table~\ref{table:thickness}. Film thicknesses for pristine samples were not measured because the ellipsometer was not housed in a controlled environment at the time of this study; removing pristine samples from their inert storage for SE characterization would have exposed them to ambient air, compromising their pristine status. However, since pristine and exposed samples were produced under identical conditions and deposition times, their film thicknesses are expected to be comparable.

\begin{table*}[!htb]
\centering
\caption{Film thickness of a representative subset of exposed tholin samples from each experimental setup, measured by SE on quartz substrates. The Mean Squared Error (MSE) is a measure of the goodness of fit for the ellipsometry model. The thickness uncertainty represents the 90\% confidence interval of the model fit. UV-generated samples were modeled as a roughness layer due to their extreme thinness. }
%Plasma samples have thicknesses of hundreds of nanometers, well above the $\sim$50--100~nm threshold required for substrate-independent surface energy measurements, whereas UV samples are sub-nanometer in thickness, well below this threshold.
\label{table:thickness}
\begin{tabular}{lcccc}
\toprule
Sample & Description & Thickness (nm) & Thickness Err. (nm) & MSE \\
\midrule
5  & JHU/PHAZER 95/5 Plasma & 941.0 & 0.2 & 2.71 \\
22 & ARC/COSmIC 95/5 Plasma & 828.6 & 0.4 & 6.73 \\
14 & JHU/PHAZER 95/5 UV     & 0.8   & 0.1 & 2.27 \\
30 & UNI/PAC 95/5 UV        & 0.2   & 0.1 & 2.12 \\
\bottomrule
\end{tabular}
\end{table*}

\section{Results and Discussion} \label{sec:results}
Contact angle measurements, summarized in Table~\ref{table3}, were used to derive the total surface energy and its corresponding dispersive and polar components for each tholin sample using the OWRK method described in Section~\ref{sec:SE}. The resulting surface energy values are summarized in Table~\ref{table4}. We first present the results examining the effects of substrate choice and air exposure on the derived surface energies (Sections~\ref{sec:substrate} and \ref{sec:exposure}). We then assess the influence of initial gas mixture composition (Section~\ref{sec:gasmix}), and compare results across the different experimental setups involved in this study (Section~\ref{sec:exp_setup}). Finally, we assess the commonality among the samples and discuss the implications of these experimental results for Titan's atmospheric and surface processes (Sections~\ref{sec:common} and \ref{sec:clouds}).

\begin{table*}[!htb]
    \centering
    \caption{Calculated total surface energies and the corresponding surface energy dispersive and polar components of all tholin samples and blank glass, quartz, and mica substrates.}
    \scriptsize
    \begin{tabularx}{\textwidth}{l|YYY}
     \toprule
        ~ & \multicolumn{3}{c}{Total Surface Energy and Dispersive and Polar Components (mJ~m$^{-2}$)} \\
        \midrule
        \multicolumn{1}{l|}{Sample} & Total Surface Energy ($\gamma_{\text{tot}}^{\text{sv}}$) & Dispersive Component ($\gamma_{\text{sv}}^{\text{d}}$) & Polar Component ($\gamma_{\text{sv}}^{\text{p}}$) \\
        \midrule
        1 (95/5 Plasma)      & 72.79~$\pm$~0.97  & 44.66~$\pm$~0.41 & 28.13~$\pm$~0.97 \\
        2 (95/5 Plasma)      & 75.55~$\pm$~0.59  & 44.26~$\pm$~0.35 & 31.29~$\pm$~0.60 \\
        3 (95/5 Plasma)      & 73.15~$\pm$~1.81  & 43.13~$\pm$~0.40 & 30.02~$\pm$~1.82 \\
        4 (95/5 Plasma)      & 68.37~$\pm$~1.59  & 37.74~$\pm$~0.31 & 30.63~$\pm$~1.59 \\
        5 (95/5 Plasma)      & 68.34~$\pm$~2.66  & 35.59~$\pm$~0.63 & 32.75~$\pm$~2.68 \\
        6 (95/5 Plasma)      & 68.42~$\pm$~2.61  & 36.63~$\pm$~0.52 & 31.79~$\pm$~2.62 \\
        7 (90/10 Plasma)     & 71.21~$\pm$~0.55  & 42.98~$\pm$~0.27 & 28.23~$\pm$~0.55 \\
        8 (98/2 Plasma)      & 70.29~$\pm$~0.74  & 36.49~$\pm$~0.35 & 33.79~$\pm$~0.77 \\
        9 (99/1 Plasma)      & 69.37~$\pm$~0.73  & 40.54~$\pm$~0.89 & 28.83~$\pm$~0.78 \\
        \midrule
        10 (95/5 UV)         & 44.09~$\pm$~1.23  & 33.44~$\pm$~0.68 & 10.65~$\pm$~1.18 \\
        11 (95/5 UV)         & 47.71~$\pm$~1.35  & 33.46~$\pm$~0.48 & 14.25~$\pm$~1.33 \\
        12 (95/5 UV)         & 35.83~$\pm$~1.02  & 27.19~$\pm$~0.77 & 8.64~$\pm$~0.94  \\
        13 (95/5 UV)         & 34.65~$\pm$~0.51  & 29.37~$\pm$~0.22 & 5.28~$\pm$~0.49  \\
        14 (95/5 UV)         & 39.69~$\pm$~1.20  & 31.36~$\pm$~1.03 & 8.33~$\pm$~1.04  \\
        15 (95/5 UV)         & 44.90~$\pm$~0.74  & 28.71~$\pm$~0.53 & 16.19~$\pm$~0.73 \\
        16 (99/1 UV)         & 53.46~$\pm$~1.10  & 36.65~$\pm$~0.45 & 16.81~$\pm$~1.09 \\
        17 (99/1 UV)         & 48.62~$\pm$~0.31  & 32.91~$\pm$~0.41 & 15.71~$\pm$~0.29 \\
        \midrule
        18 (95/5 Plasma)     & 51.27~$\pm$~1.20  & 46.81~$\pm$~0.50 & 4.46~$\pm$~1.14  \\
        19 (95/5 Plasma)     & 52.27~$\pm$~0.96  & 47.88~$\pm$~0.73 & 4.38~$\pm$~0.77  \\
        20 (95/5 Plasma)     & 51.64~$\pm$~1.60  & 49.19~$\pm$~0.29 & 2.45~$\pm$~1.58  \\
        21 (95/5 Plasma)     & 46.20~$\pm$~1.16  & 36.04~$\pm$~0.48 & 10.16~$\pm$~1.13 \\
        22 (95/5 Plasma)     & 46.21~$\pm$~1.27  & 36.11~$\pm$~0.82 & 10.09~$\pm$~1.19 \\
        23 (95/5 Plasma)     & 48.05~$\pm$~1.10  & 35.89~$\pm$~0.49 & 12.17~$\pm$~1.07 \\
        24 (90/10 Plasma)    & 36.96~$\pm$~1.03  & 34.50~$\pm$~0.94 & 2.46~$\pm$~0.70  \\
        25 (98/2 Plasma)     & 35.79~$\pm$~0.59  & 33.77~$\pm$~0.62 & 2.01~$\pm$~0.29  \\
        26 (99/1 Plasma)     & 41.37~$\pm$~0.46  & 39.19~$\pm$~0.25 & 2.18~$\pm$~0.41  \\
        27 (99/1 Plasma)     & 44.67~$\pm$~0.87  & 40.05~$\pm$~0.82 & 4.62~$\pm$~0.62  \\
        \midrule
        28 (95/5 UV)         & 64.48~$\pm$~0.65  & 31.56~$\pm$~0.54 & 32.92~$\pm$~0.72 \\
        29 (95/5 UV)         & 67.95~$\pm$~1.28  & 34.67~$\pm$~0.21 & 33.28~$\pm$~1.28 \\
        30 (95/5 UV)         & 59.24~$\pm$~1.72  & 31.70~$\pm$~0.36 & 27.53~$\pm$~1.73 \\
        31 (95/5 UV)         & 68.64~$\pm$~0.50  & 33.89~$\pm$~0.75 & 34.75~$\pm$~0.66 \\
        32 (99.9/0.1 UV)     & 59.19~$\pm$~2.30  & 31.45~$\pm$~0.41 & 27.75~$\pm$~2.31 \\
        \midrule
        Blank Glass          & 74.99~$\pm$~0.74  & 41.29~$\pm$~0.67 & 33.70~$\pm$~0.79 \\
        Blank Quartz         & 72.43~$\pm$~0.39  & 39.94~$\pm$~0.61 & 32.49~$\pm$~0.47 \\
        Blank Mica           & 77.44~$\pm$~0.55  & 42.46~$\pm$~0.74 & 34.98~$\pm$~0.64 \\
        \bottomrule
    \end{tabularx}
    \label{table4}
\end{table*}

\subsection{Effect of Substrate Choice} \label{sec:substrate}
\begin{figure}[!htb]
\centering
\includegraphics[width=\columnwidth]{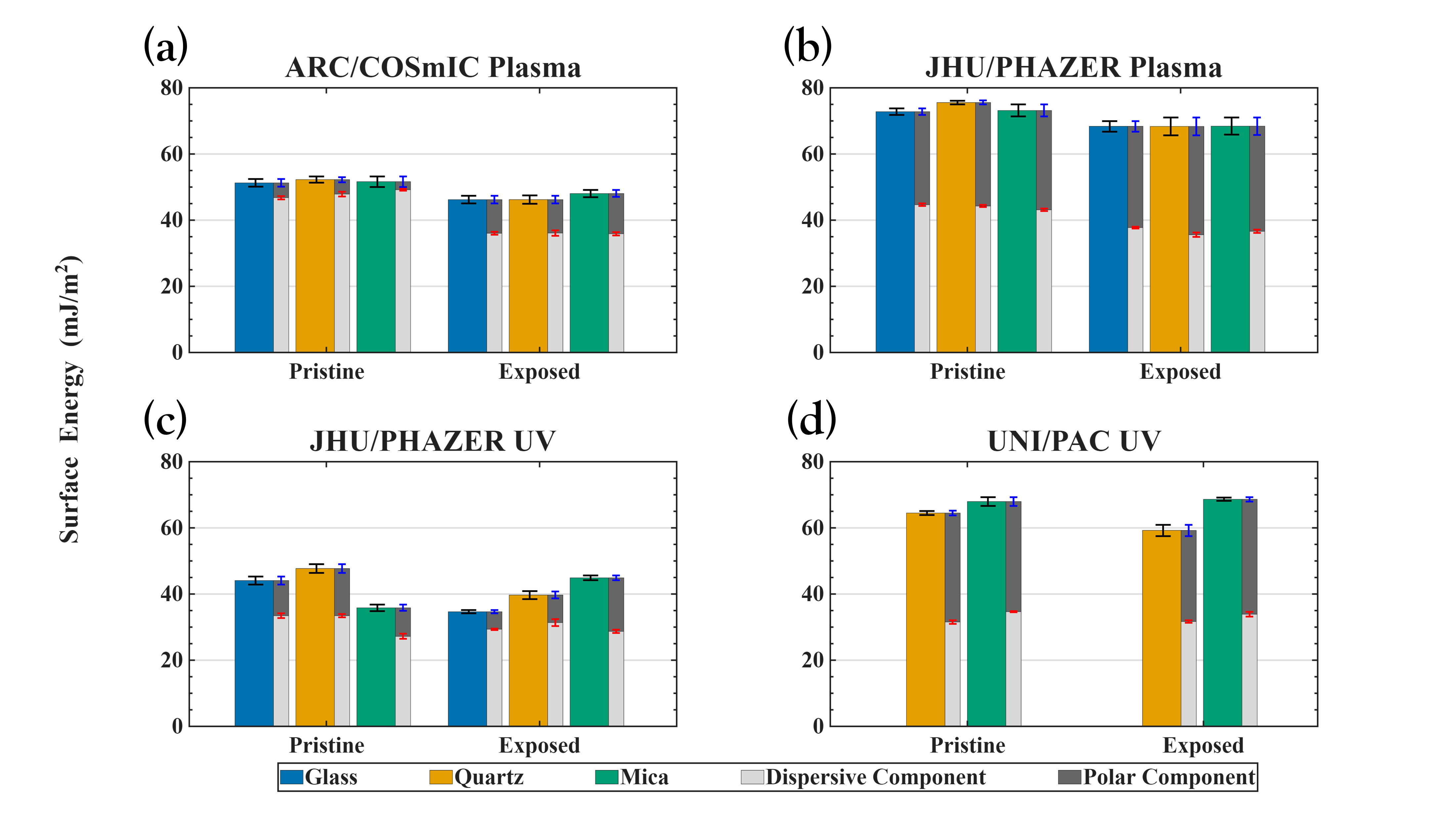}
\caption{Surface energies of pristine and air-exposed tholins produced from 5\% CH$_4$ in N$_2$ on glass, quartz, and mica: (a) ARC/COSmIC plasma, (b) JHU/PHAZER plasma, (c) JHU/PHAZER UV, and (d) UNI/PAC UV. Colored bars show total surface energy and identify the substrate; silver and gray segments show the dispersive and polar components, respectively. Black, blue, and red error bars denote 1$\sigma$ uncertainties in the total, dispersive, and polar values, respectively.}
\label{expComp}
\end{figure}

We first examine the effect of three common laboratory-grade substrates on the surface energy of 5\% CH$_4$ in N$_2$ tholin films. Comparisons among surface energies for these tholin samples deposited on glass, quartz, and mica are summarized in Figure~\ref{expComp}. 

As shown in Figure~\ref{expComp}, surface energy measurements for plasma samples (ARC/COSmIC and JHU/PHAZER plasma) exhibit only modest variation across various substrates for both the pristine and exposed samples, differing by at most 3~mJ~m$^{-2}$. The values are not statistically distinguishable in all pairwise comparisons between substrate types ($z<2.5$). Breaking surface energy into its components, the polar components are also not statistically distinguishable ($z<2.8$, absolute difference $\leq3.2$~mJ~m$^{-2}$). The dispersive components are mostly not statistically distinguishable ($z<2.7$), with two exceptions: the pristine ARC/COSmIC plasma sample on glass versus mica ($z=4.1$) and the exposed JHU/PHAZER plasma sample on glass versus quartz ($z=3.1$). Even in these cases, however, the absolute differences in $\gamma^d_{sv}$ are small ($\lesssim$2.4~mJ~m$^{-2}$). Taken together, these results indicate that the choice of substrate only introduces at most a couple of mJ~m$^{-2}$ of variation in the inferred surface energies of tholin samples made with cold plasma discharge, which is minor compared to the overall range of tholin surface energies calculated in this work.

In contrast, the UV samples exhibit a stronger dependence on substrate choice. For the JHU/PHAZER UV setup, the pristine tholin samples deposited on glass and quartz substrates exhibit higher surface energies than the tholin sample deposited on mica. While the total surface energy values derived from glass and quartz samples overlap within uncertainty ($z=2$), values for mica are lower by $7.0\sigma$ relative to quartz and $5.2\sigma$ relative to glass. The absolute differences between the samples are also much larger than for the plasma samples, with the total surface energy between the samples on the three substrates differing up to 12~mJ~m$^{-2}$. The difference in the total surface energy is mainly caused by the difference in dispersive components ($>6\sigma$ for glass-mica and quartz-mica pairs) rather than the polar components ($<3.4\sigma$ for all pairs). The exposed JHU/PHAZER UV samples show the opposite trend, with tholin deposited on mica exhibiting the highest surface energy. In this case, total surface energy values differ by $3.9\sigma$ (glass-quartz), $11.4\sigma$ (glass-mica), and $3.7\sigma$ (quartz-mica), respectively, and the largest difference is up to 10~mJ~m$^{-2}$. In this case, the polar components drive the differences between the exposed samples ($>6\sigma$ for glass-mica and quartz-mica pairs and $2.7\sigma$ for glass-quartz pairs) rather than the dispersive components ($<2.3\sigma$ for all pairs).

For the UNI/PAC setup, surface energy measurements were only obtained on two substrates due to the limited surface area of the collection chamber. For the pristine samples, surface energy values derived from contact angles measured on mica and quartz overlap within uncertainty ($z=2.4$, absolute difference 3.6~mJ~m$^{-2}$). The dispersive component shows a distinct difference between the two samples ($z=5.4$, absolute difference 3.1~mJ~m$^{-2}$), with only small differences in the polar components ($z=0.2$, absolute difference 0.4~mJ~m$^{-2}$). For the exposed samples, tholin deposited on mica exhibits a higher surface energy than that deposited on quartz by $5.2\sigma$ with an absolute difference of 9.4~mJ~m$^{-2}$, driven mainly by the difference in polar component ($3.9\sigma$ and 7.2~mJ~m$^{-2}$ in absolute difference). Overall, we do not observe any consistent trend in surface energy with substrate for the UV samples, indicating that the apparent substrate effects are more complex than for the plasma samples.

Rather than suggesting an inherent characteristic of UV samples, we attribute this behavior to the extremely low thickness of the UV samples analyzed here. Although surface energy measurements can, in principle, be performed on films consisting of only a few molecular layers \citep{Israelachvili2011}, practical contact-angle measurements require a sufficiently thick and homogeneous film to minimize thickness-dependent and substrate-induced effects. Previous studies indicate that polymer film thicknesses on the order of tens of nanometers ($\sim$50--100~nm) are typically required for surface energy to become independent of thickness and substrate choice \citep[e.g.,][]{tavana2005contact, Abbasian+2004, li2007contact,  Petek+2022}. 

Spectroscopic ellipsometry measurements (Table~\ref{table:thickness}) show that the 5\% CH$_4$ in N$_2$ plasma samples have thicknesses ranging from hundreds of nm to about one $\mu$m, whereas the 5\% UV samples are sub-nm in thickness. Although a detailed thickness-dependence study is beyond the scope of this work, ultrathin films are known to exhibit thickness-dependent surface energies until a critical thickness is reached, at which point the behavior becomes bulk-like. Thus, at ultrathin coverages, such as in the UV samples studied here, tholins are likely to nucleate in a patchy, island-like morphology (see \citep{He+2018}), resulting in discontinuous coverage that influences contact angle and surface energy measurements. As film growth proceeds and more tholin is deposited, continuous coverage develops, and the surface energy converges toward a bulk-like value independent of substrate \citep{Hwang+2015}, as observed for the plasma samples. We note, however, that the derived surface energies of all the UV samples ($35-48$~mJ~m$^{-2}$ for the JHU/PHAZER UV samples and $59-69$~mJ~m$^{-2}$ for the UNI/PAC UV samples) are statistically distinguishable from those of the same bare substrates ($z>6.8$), which all exceed 72~mJ~m$^{-2}$ (see Table~\ref{table4}). This indicates that, despite their ultrathin nature, the UV samples contribute measurably to the observed surface energy, and we are not simply probing the bare substrate. Although we cannot conclusively determine the intrinsic surface energy of the UV samples, the measurements reflect the presence of a surface layer with a lower effective surface energy than the underlying substrate.

We also compared the surface energies of a few representative 1\% CH$_4$ in N$_2$ plasma samples produced with the ARC/COSmIC and JHU/PHAZER facilities and confirmed that minimal substrate dependence holds for these plasma-generated samples. We therefore conclude that substrate choice has only a minor influence on the inferred surface energies of tholin samples once the film thickness exceeds 50--100~nm, as is the case for all plasma samples studied here, independent of initial mixtures. In contrast, surface energy values for the UV tholin samples should be interpreted with caution, as they are likely influenced by substrate effects and may not reflect the material's intrinsic bulk surface energy. As a consequence, all samples in the subsequent gas mixture comparison were produced on glass slide substrates, the most economical choice. 

\subsection{Effect of Air Exposure} \label{sec:exposure}

The second set of tests we conducted in this comparative study examined the effect of air exposure on the derived surface energy of tholin samples produced using the same 5\% CH$_4$ in N$_2$ gas mixture. Pristine samples were never exposed to ambient air as they were collected, transported, and measured under vacuum or an inert N$_2$ environment. In contrast, exposed samples were collected and transported under ambient conditions and remained exposed to laboratory air for approximately 45 days prior to characterization. Surface energy results for both pristine and exposed samples are summarized in Figure~\ref{expComp}. 

For the JHU/PHAZER plasma samples, when we compare the pristine and the exposed samples on the same substrate, the total surface energy values are not statistically distinguishable ($z<2.7$), with absolute differences of $4.4-7.2$~mJ~m$^{-2}$. The ARC/COSmIC plasma samples show slight differences between pristine and exposed samples: $z=3.0$ and $3.8$ for the glass and quartz substrates and $z=1.8$ for mica, with absolute differences up to 6~mJ~m$^{-2}$. More pronounced differences, however, are observed in the individual surface energy components. For the JHU/PHAZER plasma samples, the dispersive component decreases substantially after air exposure ($9.9-13.5\sigma$, absolute changes up to 8.7~mJ~m$^{-2}$), while the polar components remain nearly unchanged ($z<1.3$, absolute changes up to 2.5~mJ~m$^{-2}$). For the ARC/COSmIC plasma samples, both components change significantly after air exposure: the dispersive component decreases significantly ($10.7-23.4\sigma$, absolute changes up to 13.3~mJ~m$^{-2}$), whereas the polar component increases ($3.6-5.1\sigma$, absolute changes up to 9.7~mJ~m$^{-2}$).

These results indicate that although the total surface energy of the samples remains nearly unchanged, the balance between dispersive and polar interactions is substantially altered after air exposure. The changes in surface energy components are consistent with chemical modification of the near-surface region upon exposure to ambient air. Previous studies have demonstrated that tholins are highly reactive with atmospheric oxygen and water vapor. For instance, X-ray photoelectron spectroscopy depth profiling revealed that air-exposed tholin films exhibit significant oxygen contamination (up to 10~\% atomic concentration) confined to the top $\sim20$~nm of the surface \citep{Carrasco+2016}. Furthermore, hydrolysis reactions of tholins in the presence of water are known to rapidly produce oxygenated functional groups, such as carbonyls and hydroxyls, while altering existing nitrogenous structures \citep{Neish+2009}. Therefore, the increase in the polar component for the exposed ARC/COSmIC plasma sample is consistent with the incorporation of these oxygen- and water-derived functional groups, which enhance dipole-dipole and hydrogen-bonding interactions.

We note that the methods used here cannot study time-resolved exposure because the contact angle measurements on tholin samples are considered destructive -- water droplets partially dissolve the tholin films and leave ring-like residue, preventing repeated measurements on the same sample. However, an additional comparison can be made to the ARC/COSmIC sample reported in \citet{Li2022}, produced with the same 5\% CH$_4$ in N$_2$ gas mixture and stored in a laboratory refrigerator for several years prior to characterization. Although that sample was deposited on a silicon wafer substrate that is not included in the present study, our earlier results demonstrate that substrate choice has minimal influence on the surface energy of ARC/COSmIC plasma samples, allowing us to still conduct a qualitative comparison. 

The sample from \citet{Li2022} exhibits a significantly higher total surface energy than both the pristine and the 45-day exposed samples in this work, by approximately $6\sigma$ and $8\sigma$, respectively (absolute differences are 13-15~mJ~m$^{-2}$ and 18-20~mJ~m$^{-2}$). This increase is primarily driven by a substantial rise in the polar component, which reaches $\sim20$~mJ~m$^{-2}$, compared to 10-12~mJ~m$^{-2}$ for the exposed samples and 2-4.5~mJ~m$^{-2}$ for the pristine samples in this work, and is statistically distinguishable from both (by $\sim7\sigma$ and $\sim4\sigma$ level, respectively). The polar component results suggest progressive oxidation and/or hydration of the film surface with increasing exposure time to ambient conditions. On the contrary, the evolution of the dispersive component is less systematic. The 45-day exposed sample shows a decrease in the dispersive component relative to the pristine samples, where this long-term exposed sample from \citet{Li2022} exhibits a higher dispersive component than that of the 45-day exposed samples ($z\approx4.5$, absolute difference $\sim10$~mJ~m$^{-2}$) and becomes comparable to that of the pristine samples ($z<1.7$, absolute difference up to 3.5~mJ~m$^{-2}$). This non-monotonic behavior suggests that the changes in van der Waals interactions during air exposure are more complex and cannot be captured by a simple temporal trend. Future work on a time-resolved surface energy study is therefore needed to determine the evolution of dispersive interactions with exposure to ambient conditions.

The UV samples exhibit a less systematic response to air exposure, with inconsistent changes in surface energy across different substrates and setups. As established in Section~\ref{sec:substrate}, this is attributed to the ultrathin nature of the films, which makes the measurements highly sensitive to the underlying substrate and likely masks the true effect of air exposure.

Taken together, these results underscore the importance of conducting surface property measurements of tholin samples under inert conditions, as there is minimal oxygen and water vapor in Titan's atmosphere. To obtain the most physically meaningful and Titan-relevant surface energy values, samples should be stored, transported, and handled with minimal exposure to ambient air. For this reason, the remainder of this comparative study focuses exclusively on the results of the pristine tholin samples.

\begin{figure}[!htb]
\centering
\includegraphics[width=\columnwidth]{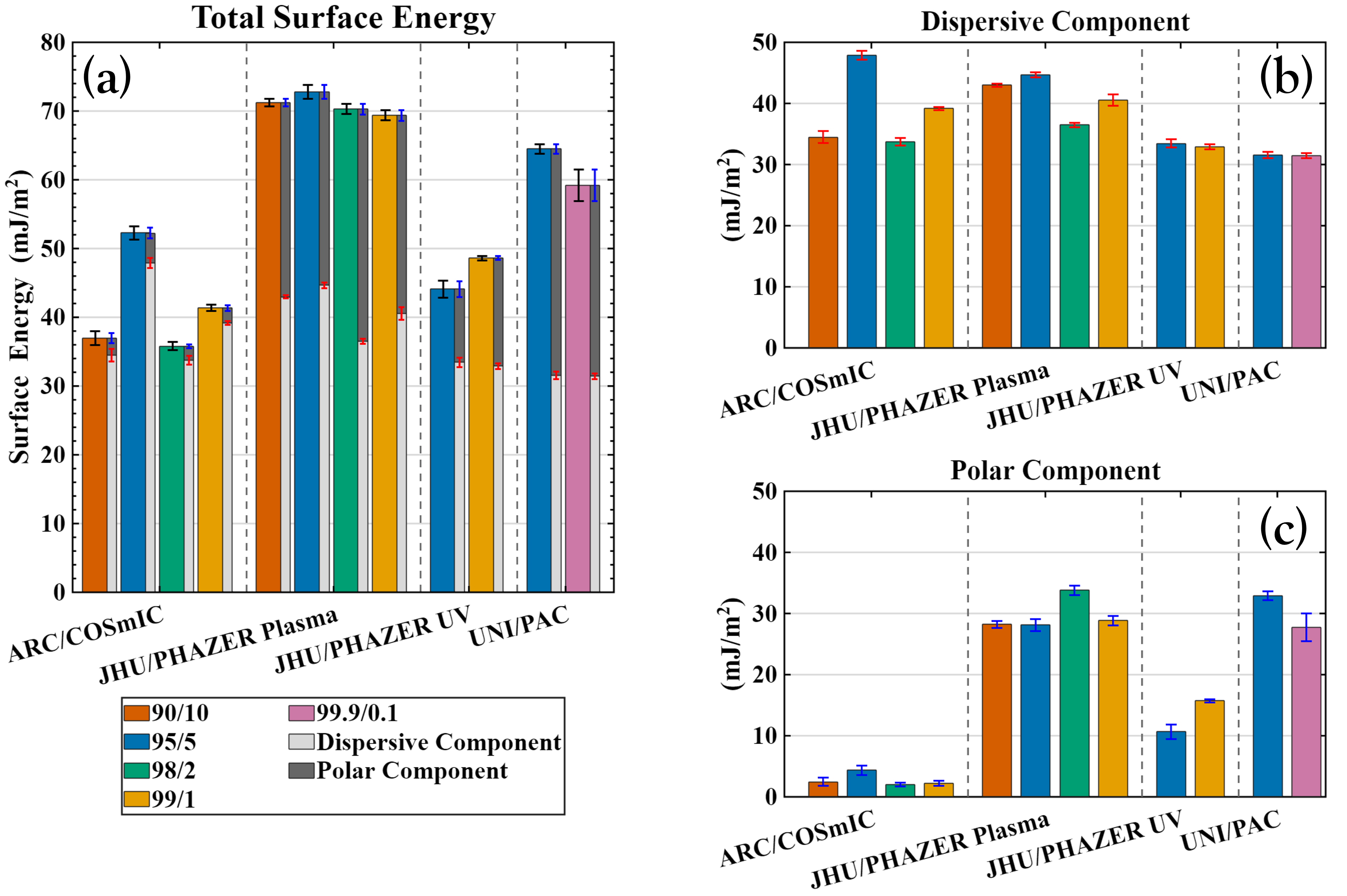}
\caption{Surface energies of pristine tholins for each gas mixture and experimental setup. Panel (a) shows total surface energy with dispersive (silver) and polar (gray) components; panels (b) and (c) emphasize the dispersive and polar components, respectively. Bar colors denote the initial N$_2$/CH$_4$ gas mixture. Black, red, and blue error bars denote 1$\sigma$ uncertainties in the total, dispersive, and polar values, respectively.}
\label{fig:gasMix}
\end{figure}

\subsection{Effect of Initial Gas Mixture} \label{sec:gasmix}
The effect of the initial N$_2$/CH$_4$ gas mixture on the surface energy of tholin samples is shown in Figure~\ref{fig:gasMix}. Based on the conclusions of the substrate and air-exposure tests presented in Sections~\ref{sec:substrate} and \ref{sec:exposure}, which demonstrate that (i) the choice of substrate minimally affects the samples once the film thickness exceeds 50--100~nm and (ii) air exposure significantly alters the calculated surface energy and its components, we restrict this comparison to pristine samples deposited on glass slide substrates, with the exception of the 5\% CH$_4$ in N$_2$ UNI/PAC sample, which was deposited on quartz.

We first examine the results of the plasma samples, as all plasma samples exceed the thickness threshold of 50--100~nm required to eliminate substrate effects. For the JHU/PHAZER plasma setup, varying the initial CH$_4$ concentration between 1--10\% results in minimal changes in the total surface energy of the samples. The total surface energy values vary less than 5\% across the full sets of samples and are not statistically distinguishable when accounting for the measurement uncertainties ($z<2.9$, absolute difference $\leq3.4$~mJ~m$^{-2}$). In contrast, the ARC/COSmIC plasma samples show a more pronounced dependence on the initial gas mixture composition, with total surface energy variations up to 30\% and statistically significant differences reaching 11.6$\sigma$ (absolute difference 15.5~mJ~m$^{-2}$). Despite this larger variation, the total surface energies of all ARC/COSmIC plasma samples remain confined to a relatively narrow absolute range (36-51~mJ~m$^{-2}$). Both plasma setups exhibit a similar qualitative pattern, with the 5\% CH$_4$ in N$_2$ gas mixture sample yielding the highest total surface energy (even though this trend is not statistically significant for the JHU/PHAZER plasma setup).

When we examine the dispersive and polar components of the surface energy for these samples, we find that the plasma samples respond to the gas mixture primarily through changes in the dispersive component. For the JHU/PHAZER plasma samples, the dispersive component varies significantly with CH$_4$ concentration in the initial gas mixture, differing by up to $15.2\sigma$ between the 2\% (lowest) and the 5\% (highest) samples. The polar component shows minimal variations among the 1\%, 5\%, and 10\% samples ($z<0.6$, absolute differences $\leq0.7$~mJ~m$^{-2}$) and only the 2\% sample exhibits a statistically significant deviation in its polar component ($z=4.5-5.9$, absolute differences up to 5.7~mJ~m$^{-2}$). A similar pattern is observed for the ARC/COSmIC plasma samples: the dispersive component differs by up to 16$\sigma$ (absolute difference 13~mJ~m$^{-2}$) between the 2\% (lowest) and the 5\% (highest) samples, whereas the polar components are statistically indistinguishable across all plasma samples ($z<2.1$, absolute difference $\leq2.5$~mJ~m$^{-2}$). 

It is interesting to note that, for both plasma setups, (i) the 5\% samples consistently exhibit the highest dispersive component across both plasma setups, (ii) the 2\% samples have the lowest dispersive component, and (iii) variations in the polar components between different initial methane concentrations are less significant than the dispersive components. This suggests that variations in methane concentration in the initial gas mixture result in a non-monotonic change in the surface energy components for the two setups since the extrema occur at 2\% and 5\% rather than at the 1\% or 10\% end-member cases. These consistent patterns across two independent plasma setups suggest that surface energy, and therefore N$_2$/CH$_4$ plasma chemistry, does not vary linearly with methane concentration. Despite operating under different pressures, plasma powers/types, and flow conditions, the two plasma setups may have similar chemical processes at play, resulting in a similar pattern observed across the samples.

This non-monotonic response to methane concentration differs from some previous studies showing systematic compositional trends. \citet{McDonald+1994} and \citet{Horst+2018} found that lower CH$_4$ mixing ratios lead to more nitrogen-rich tholins and gas-phase products, respectively. \citet{Mahjoub+2012} demonstrated that optical constants vary systematically with methane concentration. However, nonlinear behavior has been observed in other tholin properties. \citet{Sciamma-OBrien2010-qi} identified two competitive chemical regimes in tholin production, with a threshold around 5\% CH$_4$ in N$_2$ where an inhibiting process begins to dominate over the growth process. \citet{Dubois2020-dw} found that different CH$_4$ concentrations produce distinct ion populations, with amine-rich chemistry at low concentrations and aliphatic-rich chemistry at higher concentrations. The non-monotonic surface energy pattern observed here, with extrema at intermediate concentrations, suggests that competing chemical pathways rather than simple compositional scaling govern surface properties.

For instance, the observed non-monotonic dependence of surface energy on the initial CH$_4$ concentration, exhibiting a peak in the dispersive component at 5\% and a trough at 2\%, may reflect fundamental shifts in the macromolecular structure of the tholins. Previous structural characterizations have demonstrated that the N$_2$/CH$_4$ ratio dictates the balance between aliphatic and aromatic or nitrogenated structures: low CH$_4$ concentrations (1-2\%) tend to produce amine-rich polymers with high nitrogen incorporation, whereas higher CH$_4$ concentrations yield increasingly aliphatic, hydrocarbon-rich structures \citep{Gautier2012-ix, Dubois2020-dw}. Furthermore, the 5~\% CH$_4$ concentration has been identified as an optimal regime for gas-to-solid conversion, maximizing the production of C$_2$ intermediates that drive efficient copolymerization \citep{Sciamma-OBrien2010-qi, Gautier+2014}. As suggested by \citet{Cable2012-sf}, such compositional transitions are indicative of underlying shifts in the sp$^2$/sp$^3$ carbon ratio. We hypothesize that the peak in surface energy at 5\% CH$_4$ is a macroscopic manifestation of this optimal copolymerization, which may maximize structural complexity, cross-linking, and the density of polar functional groups at the surface, but spectroscopy (e.g., Raman, solid-state NMR) of these specific samples would be required to definitively correlate the surface energy with the sp$^2$/sp$^3$ ratio and aromatic cluster size. Moreover, we hypothesize that the dispersive component of the surface energy is a macroscopic indicator of the efficiency of this molecular growth. Because dispersive interactions scale with molecular polarizability (which increases with the size and cross-linking of the carbon backbone) the peak in the dispersive component at 5~\% CH$_4$ likely reflects the formation of larger, more complex macromolecular structures. Conversely, the polar component, which remains relatively consistent across gas mixtures, may primarily account for the terminal nitrogen-bearing functional groups (e.g., amines and nitriles) that cap these polymer chains.

Our interpretation of the UV samples is more limited due to their ultrathin nature. As discussed in Section~\ref{sec:substrate}, both UV setups with the 5\% CH$_4$ in N$_2$ gas mixture produce films that are sub-nm in thickness, well below the 50--100~nm threshold required to suppress substrate effects. Instead of producing the rest of the samples following the plasma gas-mixture sequence (1\%, 2\%, 10\% CH$_4$ in N$_2$), we selected methane concentrations that maximize sample yield for both UV setups. Because CH$_4$ is optically absorbing in the wavelength range of the UV lamps employed for both setups ($115-400$~nm), reducing the CH$_4$ concentration increases the photon penetration depth and thus enhances the production efficiency of the samples. Accounting for the different operating pressures of the two UV setups (2~Torr for JHU/PHAZER and 500~Torr for UNI/PAC), we choose to use a 1\% mixture for the JHU/PHAZER setup and a 0.1\% mixture for the UNI/PAC setup, so the number densities of methane are comparable between the two setups. Unfortunately, even with the low methane concentration and extended experimental run times (144~hr for JHU/PHAZER and 421~hr for UNI/PAC, see Table~\ref{table1}), the resulting UV samples reach thicknesses of only $2-3$~nm, which is still well below the thickness threshold of 50--100~nm. Therefore, comparisons between UV samples should be interpreted with caution.

The UV samples show an inconsistent response to changes in the initial gas mixture. For example, the JHU/PHAZER UV samples show a statistically significant difference between the 1\% and 5\% CH$_4$ mixtures on glass, but not on quartz. As established in Section~\ref{sec:substrate}, this inconsistent behavior is attributed to the ultrathin nature of the films, which makes the measurements highly sensitive to the underlying substrate and prevents a reliable assessment of the true effect of the gas mixture.

Overall, while the magnitude of concentration dependence on total surface energy differs between experimental setups -- being minimal for the JHU/PHAZER plasma samples and more pronounced for the ARC/COSmIC samples -- the total surface energy remains confined to a relatively narrow range for each respective setup across all methane concentrations explored. In the context of Titan's atmosphere, this suggests that local variations in methane abundance are unlikely to strongly alter the bulk adhesion/cohesion behaviors (i.e., the ``stickiness") of the haze particles. In contrast, the individual surface energy components, especially the dispersive component, show non-linear variations with methane concentration for both plasma setups, hinting at some levels of chemical changes between the samples that affect non-polar interactions. However, the polar components remain relatively constant across gas mixtures, indicating a similar degree of nitrogen incorporation into the samples, resulting in a similar extent of dipole-dipole and hydrogen-bonding interactions.

\subsection{Effect of Experimental Setup} \label{sec:exp_setup}

We finally examine how the choice of experimental setup influences the surface energy of tholin samples. Figure~\ref{fig:gasMix} summarizes the surface energy results for pristine samples produced in the three laboratories across the range of N$_2$/CH$_4$ gas mixtures used in this study. There are substantial differences in the resulting surface energy between the two plasma experimental setups (JHU/PHAZER and ARC/COSmIC). Although both plasma setups exhibit similar qualitative trends in surface energy as a function of methane concentration, samples produced with the same gas mixture show large, statistically significant differences between the two laboratories. The JHU/PHAZER plasma samples exhibit the highest surface energies, ranging from 69.4 to 75.6~mJ~m$^{-2}$, with the least internal variation compared to other labs. In contrast, the total surface energies of the ARC/COSmIC plasma samples span a lower range, from 35.8 to 51.3~mJ~m$^{-2}$. Even the JHU/PHAZER plasma sample with the lowest surface energy exceeds the ARC/COSmIC plasma sample with the highest surface energy by 18.1~mJ~m$^{-2}$ ($12.9\sigma$), and when comparing plasma samples produced with the same gas mixture, the differences between the two plasma setups become even more pronounced, reaching 32.5$\sigma$, 36.5$\sigma$, 13.9$\sigma$, and 29.3$\sigma$ for the 1\%, 2\%, 5\%, and 10\% CH$_4$ in N$_2$ mixtures, respectively. These differences are substantially larger than the gas mixture-induced variations observed within a single setup. When we examine the composing components of the surface energy, we find that the disparity in total surface energy arises almost entirely from differences in the polar component. For the 1\%, 2\%, and 5\% CH$_4$ in N$_2$ samples, the dispersive components have only modest differences between the two setups with $z=1.5-3.8$ and absolute differences $\leq3$~mJ~m$^{-2}$, whereas the polar components differ dramatically (16-39$\sigma$ for the four gas mixtures, absolute differences between $24-32$~mJ~m$^{-2}$). Only the 10\% sample shows a more significant difference in dispersive component ($8.7\sigma$, absolute difference 8.5~mJ~m$^{-2}$).

The large difference observed in the polar components suggests that the dominant factor distinguishing the two plasma setups is their relative efficiency in dissociating nitrogen and generating polar, nitrogen-bearing functional groups (\citealt{Li2022}; The specific discharge parameters for each plasma setup are summarized in Table~\ref{table1}; detailed descriptions of the JHU/PHAZER AC glow discharge and the ARC/COSmIC pulsed DC discharge can be found in \citet{Sebree2018-zy} and \citet{Sciamma-OBrien2014-fg}, respectively.). We observe that samples produced using the JHU/PHAZER plasma setup exhibit the highest total surface energies (up to 72.79~mJ~m$^{-2}$), driven by elevated polar components (up to 33.79~mJ~m$^{-2}$). On the other hand, the ARC/COSmIC plasma samples show some of the lowest total energies (down to 35.79~mJ~m$^{-2}$), corresponding to much smaller polar contributions (as low as 2.01~mJ~m$^{-2}$). 

Interestingly, bulk elemental composition alone does not readily explain this difference. Previous compositional analyses \citep{Nuevo2022-bg, He2022} show that the C/N ratios for the solid products are comparable between the two setups, and for the 10\% CH$_4$ sample, the ARC/COSmIC samples even exhibit a lower C/N ratio than the JHU/PHAZER samples (0.9 versus 1.5). This suggests that the differences in the polar component of surface energy are unlikely to be controlled by total nitrogen content of the samples, but rather by the chemical form in which nitrogen is incorporated, specifically, the fraction incorporated into highly polar functional groups at the surface.

Infrared spectroscopy studies have identified primary amines ($-$NH$_2$) and nitrile groups ($-$C$\equiv$N) as the dominant nitrogen-containing functional groups of tholins made with cold plasma \citep{Imanaka2004-nt, He2012, Derenne2012-jm, Gautier+2014}. Of these, primary amines are substantially more polar: they act as both hydrogen-bond donors and acceptors, whereas nitrile groups are only hydrogen-bond acceptors, and nitrogen incorporated into aromatic ring networks contributes minimally to surface polarity due to delocalization of the lone-pair electrons \citep{Imanaka2004-nt}. Quantitative measurements show that plasma (spark discharge) samples contain 2--3 times more primary amines than UV samples \citep{cable2014identification}, consistent with the higher polar surface energy components we observe for plasma samples. Furthermore, \citet{Imanaka2004-nt} demonstrated that the form of nitrogen incorporation changes with synthesis conditions: at higher pressures, nitrogen tends to be incorporated as terminal nitrile groups, while at lower pressures it is more likely incorporated into carbon networks as carbodiimides or isocyanides, both of which contribute less to surface polarity than free amine groups. These findings suggest that the large differences in polar component between JHU/PHAZER and ARC/COSmIC plasma samples could be due to differences in the relative abundance of amine versus nitrile functional groups at the tholin surface, driven by differences in plasma conditions such as pressure and gas exposure time.

Two main experimental factors may contribute to the differences in the magnitudes of the polar components between the two setups. The JHU/PHAZER setup employs a continuous AC plasma, which has been shown to generate a higher density of reactive excited species and has a higher energy utilization efficiency than DC or pulsed plasma sources \citep{Liu2013-gz}. In contrast, the pulsed plasma energy source in COSmIC tends to favor the production of simpler chemical products \citep{Sciamma-OBrien2014-fg}. In addition, the gas exposure time differs significantly between the two setups: approximately 3~s for the JHU/PHAZER setup compared to 3.5~$\mu$s for the ARC/COSmIC setup (six orders of magnitude shorter). The substantially longer exposure time in the JHU/PHAZER setup allows significantly more time for complex reactions to occur. In fact, previous chemical characterization of ARC/COSmIC samples indicates that only the first steps of the chemical chain are occurring for the N$_2$/CH$_4$ gas mixture, and additional precursor molecules such as acetylene (C$_2$H$_2$) and benzene (C$_6$H$_6$) have to be introduced to promote more complex chemistry \citep{Sciamma-OBrien2014-fg}. Taken together, these factors suggest that the JHU/PHAZER plasma setup, with its continuous energy input and longer gas exposure time, may provide a more favorable environment for the formation of polar functional groups in the solid tholin samples.

In the context of Titan's atmosphere, these two plasma setups may represent different stages of haze evolution at different altitudes. The ARC/COSmIC setup, with its short gas exposure time and pulsed energy source, is designed to simulate the initial steps of the chemical chain (the primary dissociation of N$_2$ and CH$_4$) that occur high in Titan's ionosphere and thermosphere (above $\sim900$~km) where energetic electrons initiate the chemistry \citep{Sciamma-OBrien2014-fg}. In contrast, the JHU/PHAZER setup, with its continuous energy input and longer residence time, allows for extensive secondary chemistry and molecular growth. This environment may be more representative of the mature haze particles found at lower altitudes in the stratosphere, where particles have had sufficient time to undergo complex reactions and accumulate polar functional groups before eventually settling to the surface \citep{He2017}.

For the UV samples, a consistent difference is observed between the two experimental setups. For the same substrate, the UNI/PAC samples consistently exhibit higher total surface energies than the JHU/PHAZER samples ($z>11.2$, absolute difference $>17$~mJ~m$^{-2}$), a difference driven primarily by a larger polar component ($z>12.3$, absolute difference $>19$~mJ~m$^{-2}$). While this comparison must be interpreted with caution due to the substrate effects established in Section~\ref{sec:substrate}, the trend holds across different substrates, suggesting it may reflect an intrinsic difference between the two setups. We also note that the thicker UV samples produced at lower methane concentrations in both setups exhibit similar surface energy values that do not trend further toward substrate-dominated behavior. This is another indication that the elevated surface energy of the UNI/PAC UV samples may reflect a genuine larger polar component contribution, possibly related to the 60-fold longer gas exposure time in the UNI/PAC chamber. 

We lastly compare the JHU/PHAZER plasma and UV samples, which offer a controlled assessment of the effect of energy source, as all other experimental conditions are identical. Even though the absolute surface energy values of the UV samples are influenced by substrate effects, both JHU/PHAZER UV samples are statistically distinguishable from their bare substrates, exhibiting significantly lower total surface energies ($>16\sigma$, absolute difference $>21$~mJ~m$^{-2}$) driven mainly by the reduced polar components ($>12\sigma$). This demonstrates that the low polar component observed in the UV samples (between 8.64-16.81~mJ~m$^{-2}$ compared to 32.49-34.98~mJ~m$^{-2}$ for the bare substrates) is a real material property rather than an artifact of probing the underlying substrate, even though the exact value may be uncertain due to the ultrathin film thickness. 

Having established that the surface energy of the JHU/PHAZER UV samples is statistically distinguishable from their bare substrates, a direct comparison with the corresponding plasma samples shows a systematic difference between the two energy sources. For both the 1\% and 5\% CH$_4$ samples, the JHU/PHAZER plasma samples consistently exhibit higher total surface energy than the UV samples ($>12\sigma$, absolute difference $>15$~mJ~m$^{-2}$), driven by substantially higher polar component ($>9\sigma$, absolute difference $>12$~mJ~m$^{-2}$). This indicates that nitrogen incorporation into polar functional groups is more efficient for the plasma energy source. This behavior is expected, as the wavelength range of the UV lamp ($115-400$~nm) cannot directly dissociate N$_2$ and nitrogen can only be incorporated through secondary pathways \citep{Trainer2012-rc}, whereas the continuous AC plasma source effectively breaks the nitrogen triple-bond directly and readily promotes the formation of nitrogen-bearing functional groups in the solid tholin samples \citep{Horst+2018}.

\subsection{Common Traits of Tholin Samples} \label{sec:common}
Across all pristine tholin samples measured in this study, the total surface energy varies between 36~mJ~m$^{-2}$ (ARC/COSmIC, 2\% CH$_4$ in N$_2$, on glass) and 76~mJ~m$^{-2}$ (JHU/PHAZER plasma, 5\% CH$_4$ in N$_2$, on quartz), yielding a total range of 40~mJ~m$^{-2}$. This large range is dominated by variation in the polar component, which ranges between 2~mJ~m$^{-2}$ (ARC/COSmIC, 2\% CH$_4$ in N$_2$, on glass) and 34~mJ~m$^{-2}$ (JHU/PHAZER plasma, 2\% CH$_4$ in N$_2$, on glass), a range of 32~mJ~m$^{-2}$. The dispersive components, on the other hand, vary over a narrower interval, between 27~mJ~m$^{-2}$ (JHU/PHAZER UV, 5\% CH$_4$ in N$_2$, on mica) and 45~mJ~m$^{-2}$ (JHU/PHAZER plasma, 5\% CH$_4$ in N$_2$, on glass), corresponding to a maximum spread of 18~mJ~m$^{-2}$. Restricting the analysis to plasma samples, whose surface energy values are less susceptible to substrate effects, does not change this overall picture. The variation in total surface energy still spans nearly 39~mJ~m$^{-2}$, which is almost entirely due to the 32~mJ~m$^{-2}$ spread in the polar component. The dispersive components show much smaller variance (34-45~mJ~m$^{-2}$), a range of only 11~mJ~m$^{-2}$. 

Thus, even with an expanded range of initial gas compositions, our results confirm the conclusion of \citet{Li2022} that tholins exhibit consistently high surface energies dominated by the high dispersive components, with much of the inter-sample variability arising from the polar components. To place these values in context, the surface energies of tholins are higher than those of common polymers, which typically range between 20 and 50~mJ~m$^{-2}$ \citep{Owens1969, Wu+1971, Wu1982, Van_Oss2006-sq}. Instead, tholin surface energies are closer to that of amorphous carbon ($59\pm3$~mJ~m$^{-2}$), as measured by the same contact angle method \citep{McGuiggan+2002}. This general behavior suggests that if Titan's hazes are similar to tholins, they should also possess a high surface energy with an elevated dispersive component. 

The high dispersive component observed across all tholin samples is consistent with their high degree of unsaturation. Solid-state $^{13}$C NMR studies of plasma tholins indicate that approximately 60\% of carbon atoms are in sp$^2$ configurations, including aromatic, heteroaromatic, and olefinic carbons \citep{Derenne2012-jm}, and the delocalized $\pi$ electrons of these structures are highly polarizable, giving rise to strong London dispersion interactions \citep{Fowkes1964-wj}. The broadly similar dispersive components across setups therefore reflect the shared sp$^2$-rich carbon backbone common to all laboratory-produced tholins. The modest variation in the dispersive component that does exist between setups likely reflects differences in the degree of aromatic condensation. \citet{Imanaka2004-nt} demonstrated using IR, UV/VIS, and Raman spectroscopy that tholins produced at lower pressures contain more abundant nitrogen-containing polycyclic aromatic compounds with larger fused ring clusters and greater delocalized $\pi$ electron density, while higher-pressure tholins have more aliphatic, saturated structures with smaller aromatic cluster sizes. Since the dispersive component of surface energy scales with molecular polarizability \citep{Fowkes1964-wj}, and polarizability increases with aromatic ring size and conjugation length, differences in aromatic condensation between setups would produce the modest but measurable differences in dispersive component that we observe. This is further supported by the finding that optical constants of tholins, which are directly related to molecular polarizability, vary with CH$_4$ concentration \citep{Mahjoub+2012}, consistent with the weak but non-zero dependence of dispersive component on gas mixture observed here.

On the other hand, the highly varying polar components between experimental setups indicate that nitrogen incorporation is complex and highly dependent on the experimental parameters, especially energy source and gas exposure time. For Titan, this implies that the polar components of haze particles likely depend on the altitude at which they formed and the specific energy source driving the local chemistry (e.g., solar UV photons versus magnetospheric electrons or cosmic rays). Although this variability in polar components may not significantly alter their interactions with nonpolar hydrocarbon lakes and clouds, it could influence their reactivity with atmospheric species with non-zero polar components, such as HCN \citep{Yu_2023-zd}.

\subsection{Implications Regarding Titan Processes} \label{sec:clouds}
Understanding the surface energy of Titan's haze particles has important implications for Titan's atmospheric and surface processes, especially cloud formation through heterogeneous nucleation and the interaction of haze particles with Titan's hydrocarbon lakes \citep{Yu2020}. Clouds can form from the condensation of organic molecules such as CH$_4$ and ethane (C$_2$H$_6$) in Titan's atmosphere \citep{Anderson+2018, Barth2017-uw}. While homogeneous nucleation is one possible pathway for cloud formation, heterogeneous nucleation facilitated by cloud condensation nuclei (CCN) is generally considered the dominant mechanism because it requires a lower degree of supersaturation \citep{Pruppacher1996-vp}. The abundance of organic-solid haze particles in Titan's atmosphere makes them natural candidates for CCN. Previous studies have suggested that, for haze particles to act as effective CCN for hydrocarbon clouds such as methane and ethane, the contact angle between the organics and the condensing liquid or solid hydrocarbons should fall within the range of 12-45$^\circ$ \citep{McDonald1964-wl, Mahata1975-bh, Yu2021-sa, Li2022}. As a conservative working criterion, we categorize samples with predicted contact angles less than 30$^\circ$ as good CCN.

The interaction between haze particles and Titan's lakes, primarily composed of liquid CH$_4$, C$_2$H$_6$, and N$_2$ \citep{Malaska+2017, Farnsworth+2019}, is also particularly interesting \citep{Mastrogiuseppe2016-lo, Mastrogiuseppe2018-wd, Poggiali2020-ax}. While nitrogen exists as a dissolved gas rather than a pure liquid at Titan's surface conditions, we use the surface tension of liquid N$_2$ in our calculations as a proxy to represent the nitrogen component of this ternary mixture. Haze particles may float on the surface of these lakes if they are less dense than the liquid hydrocarbons or if they form a non-zero contact angle with the lake's liquids \citep{Cordier2019-dl, Yu2020}. Previous studies have indicated that tholin densities can vary from $0.4-1.4$~g~cm$^{-3}$, with most measurements exceeding the densities of liquid hydrocarbons ($0.4-0.75$~g~cm$^{-3}$) \citep[see a summary of tholin's densities in][]{Yu_2023-zd}. Therefore, here we assume that our tholin samples are, in general, denser than hydrocarbon lake liquids, and that the likelihood of floating primarily depends on the contact angle. A non-zero contact angle would result in flotation of haze particles, potentially contributing to the observed wave-dampened surfaces of the lakes and to the transient ``magic islands" features \citep{Stephan2010, Barnes2011, Soderblom2012, Zebker2014, Grima2017, Yu2024}.

Since we cannot directly measure contact angles between liquid or solid CH$_4$, C$_2$H$_6$, or N$_2$, and tholin, we predict the contact angles between these relevant hydrocarbons and each pristine tholin sample. While other condensable species such as C$_6$H$_6$, HCN, and HC$_3$N are also known to form clouds in Titan's atmosphere \citep{Anderson+2018}, we focus our current predictions on the major cloud species CH$_4$ and C$_2$H$_6$. Surface energy components for a broader suite of Titan condensates have been reported by \citet{Yu_2023-zd}; a subsequent companion study will combine those values with the tholin surface energies measured here to evaluate nucleation and wetting for these additional cloud-forming species. Also, it is important to clarify that a contact angle between two solids is a theoretical construct, not a directly measurable quantity. It represents the thermodynamic compatibility between a solid condensate (e.g., C$_2$H$_6$ ice) and a solid substrate (tholin) \citep{Owens1970, Israelachvili2011}.

\begin{figure*}[!htb]
    \centering
    \includegraphics[width=\textwidth]{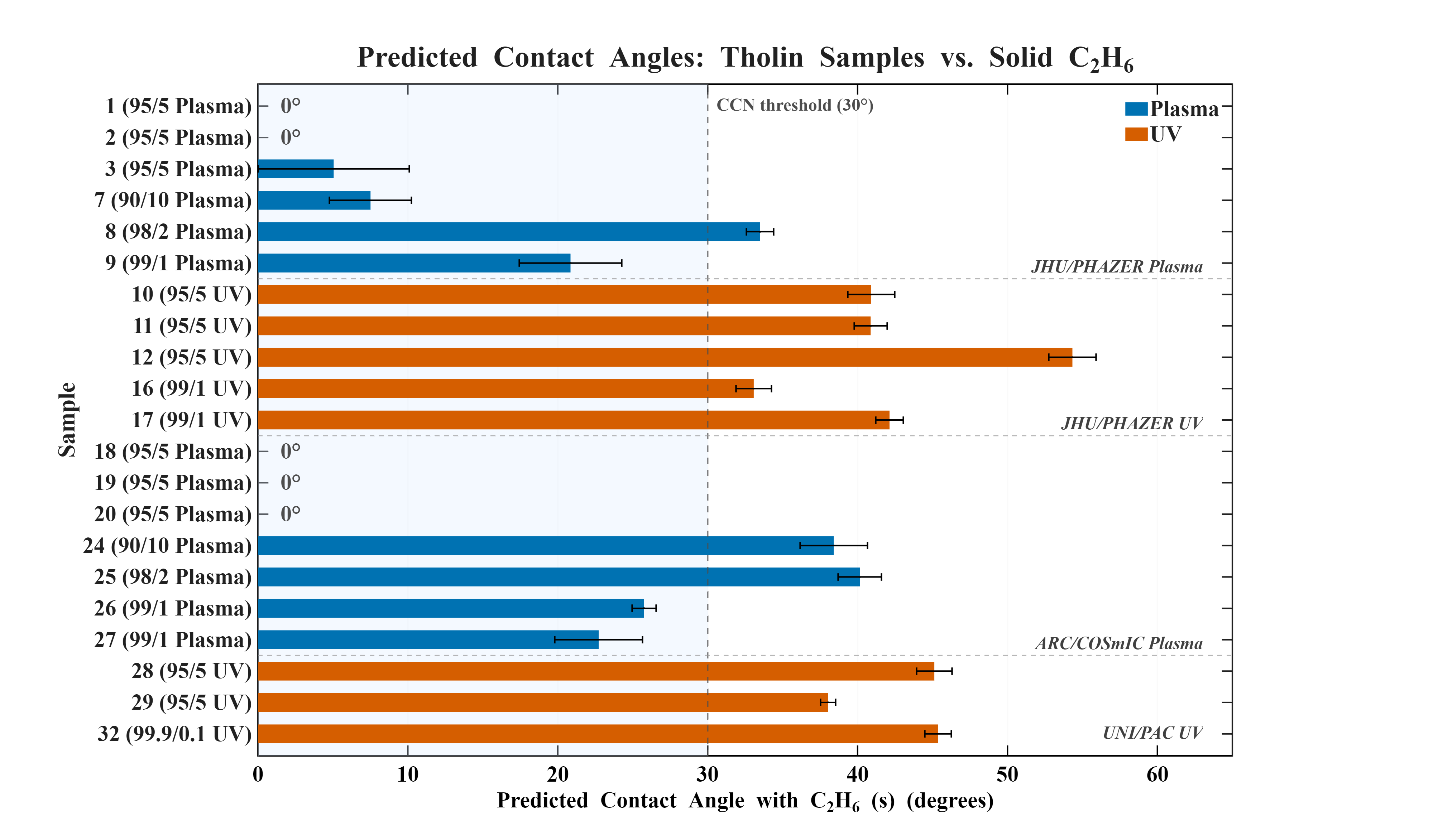}
    \caption{Predicted contact angles between each pristine tholin sample and solid C$_2$H$_6$ condensate, grouped by experimental setup. Bars are color-coded by energy source: blue for plasma discharge (JHU/PHAZER and ARC/COSmIC) and orange for UV irradiation (JHU/PHAZER and UNI/PAC). Horizontal error bars represent 1$\sigma$ uncertainties propagated from the measured surface energy components. The vertical dashed line at 30$^\circ$ marks the upper boundary of the contact angle range considered favorable for efficient cloud condensation nuclei (CCN) activity \citep{McDonald1964-wl, Mahata1975-bh, Yu2021-sa, Li2022}. The light blue shaded region ($\theta < 30^\circ$) highlights the CCN-favorable zone. Samples are grouped by experimental setup, with italic labels indicating the setup name at the bottom right of each group. Predicted contact angles with solid and liquid CH$_4$ and liquid N$_2$ are $0^\circ$ for all samples; these complete-wetting cases are therefore not shown. For liquid C$_2$H$_6$, all plasma samples also yield $\theta = 0^\circ$; non-zero predicted angles are found only for three UV samples: $31.76^\circ \pm 2.85^\circ$ (sample 12, JHU/PHAZER UV), $6.38^\circ \pm 6.38^\circ$ (sample 28, UNI/PAC UV), and $6.29^\circ \pm 6.29^\circ$ (sample 32, UNI/PAC UV).}
    \label{fig:predictedAngles}
\end{figure*}

The prediction is derived from an extension of the Young-Dupr\'e equation, where the work of adhesion between the two solids is estimated from their respective surface energy components using the geometric mean combining rule \citep{Fowkes1964-wj, Owens1969}. This theoretical angle serves as a proxy for nucleation efficiency; a low predicted angle implies favorable wetting and suggests the substrate is a good candidate for a cloud condensation nucleus. We calculate this predicted angle using the methodology described in Section~3.4 of \citet{Li2022}. This is done by rearranging Equation~\ref{eqn3} to obtain:

\begin{equation}
\label{eqn4}
\cos\theta=\frac{2(\sqrt{\gamma^{d}_{sv}\gamma^{d}_{lv}}+\sqrt{\gamma^{p}_{sv}\gamma^{p}_{lv}})}{\gamma_{lv}^{tot}}-1,
\end{equation}
where $\gamma_{lv}^{tot}$, $\gamma^{d}_{lv}$, and $\gamma^{p}_{lv}$ represent the condensate total surface free energy and its dispersive and polar components, respectively (see Table~5 of \citealt{Yu2020}), and $\gamma_{sv}^{tot}$, $\gamma^{d}_{sv}$, and $\gamma^{p}_{sv}$ represent the total surface energy and corresponding components of tholin. The surface tension values used for the condensates correspond to cryogenic temperatures relevant to Titan's surface conditions ($\sim94$~K; \citealt{Yu2020}). All cloud condensates (CH$_4$ and C$_2$H$_6$ liquids and ices) and lake liquid constituents (liquid CH$_4$, C$_2$H$_6$, and N$_2$) considered here are nonpolar, so $\gamma^{p}_{lv}=0$. Therefore, the dispersive component equals the total surface energy ($\gamma^{d}_{lv}=\gamma_{lv}^{tot}$) of the condensates. Equation~\ref{eqn4} then reduces to:
\begin{equation}
\label{eqn5}
\cos\theta=2\sqrt{\frac{\gamma^{d}_{sv}}{\gamma^{d}_{lv}}}-1.
\end{equation}
From Equation~\ref{eqn5}, if $\gamma^{d}_{sv}\ge\gamma^{d}_{lv}$, the right-hand side would exceed unity, which is unphysical and instead indicates the complete-wetting limit $\theta=0^\circ$. In other words, when the dispersive component of the surface energy of tholin exceeds that of the condensate, complete wetting is expected \citep{Fox1952-oo}. On the contrary, if tholin has a lower dispersive component than the condensate, a finite contact angle is predicted. We therefore compute:
\begin{equation}
\label{eqn6}
\theta = \left\{
\begin{array}{ll}
      0, & \text{if } \gamma^{d}_{sv}\ge\gamma^{d}_{lv} \\
      \arccos(2\sqrt{\frac{\gamma^{d}_{sv}}{\gamma^{d}_{lv}}}-1), & \text{if } \gamma^{d}_{sv}<\gamma^{d}_{lv}. \\
\end{array} 
\right.
\end{equation}
The resulting predicted contact angles are summarized in Figure~\ref{fig:predictedAngles}. 

Even though the tholin samples from the three independent laboratories differ most strongly in their polar components, only the dispersive components enter Equations~\ref{eqn5}-\ref{eqn6}. As a result, many samples behave similarly in terms of predicted contact angles. All samples display a predicted contact angle of 0$^\circ$ with CH$_4$, due to the low surface tension of liquid methane and the low surface energy of methane ice \citep{Yu_2023-zd}. Our analysis of predicted contact angles with C$_2$H$_6$ ice reveals more distinct trends among the tholin samples. For the 5\% and 1\% CH$_4$ in N$_2$ gas mixture samples, all plasma tholins exhibit predicted angles $<30^\circ$. The 2\% and 10\% JHU/PHAZER plasma samples also have predicted angles $<30^\circ$, but the ARC/COSmIC plasma samples made with these mixtures have $\theta>30^\circ$. On the other hand, all UV samples have predicted contact angles with C$_2$H$_6$ ice exceeding $30^\circ$.

To assess the physical significance of the dispersive component differences reported in Sections~\ref{sec:substrate}--\ref{sec:exp_setup}, we note that Equation~\ref{eqn6} predicts complete wetting ($\theta=0^\circ$) whenever $\gamma^d_{sv}\ge\gamma^d_{lv}$. For solid C$_2$H$_6$ ($\gamma^d_{lv}=43.4$~mJ~m$^{-2}$), all samples with $\gamma^d_{sv}\geq43.4$~mJ~m$^{-2}$ yield $\theta=0^\circ$ regardless of modest inter-sample differences. Near this threshold, a 2.4~mJ~m$^{-2}$ dispersive difference, the maximum substrate-induced variation observed for plasma samples (Section~\ref{sec:substrate}), corresponds to a change of $\sim5-8^\circ$ in predicted contact angle. However, all 5~\% CH$_4$ plasma samples remain in the CCN-favorable regime ($\theta<30^\circ$) across all substrates. Only when the dispersive component falls well below the condensate threshold, as occurs for ARC/COSmIC 2\% and 10\% samples and all UV samples, does the predicted contact angle exceed $30^\circ$ and become unfavorable for CCN activity.

These results have direct implications for cloud formation on Titan. The finding that all of our tholin samples would serve as effective CCN for CH$_4$ clouds is consistent with the abundant CH$_4$ clouds observed in Titan's atmosphere \citep{Griffith1991-oq, Griffith1998-fm, 2000Sci...290..509G, Bouchez2004-ke, 2005Sci...310..474G, 2006Natur.442..432T}. This is not the case for C$_2$H$_6$ ice clouds, however. As shown in Figure~\ref{fig:predictedAngles}, the plasma tholins are predicted to be effective CCN with ethane ice clouds, while the UV tholins are not. Given that C$_2$H$_6$ ice clouds have been observed in Titan's troposphere \citep{2006Sci...313.1620G,anderson2011titan}, our findings suggest that tholins produced by the cold-plasma discharge may be better analogs for Titan's haze particles in the context of ethane cloud nucleation. This conclusion, however, should be treated with caution, as the intrinsic surface energies of the ultrathin UV tholin samples remain uncertain.

The conclusion that plasma tholins are the most representative analogs also has implications for haze-lake interactions. Our predictions indicate that all tholin samples would be completely wetted by liquid N$_2$ and CH$_4$ ($\theta=0^\circ$). For liquid C$_2$H$_6$, which has a higher surface tension, the plasma tholins also predict complete wetting. If Titan's haze has surface properties similar to these plasma tholins, the particles would be unable to float and would instead sink into the hydrocarbon lakes.

In summary, our surface energy measurements and the resulting contact angle predictions suggest that Titan-like tholins are highly wettable by CH$_4$ and likely by C$_2$H$_6$ as well, favoring their role as efficient CCN for hydrocarbon clouds while disfavoring flotation on Titan's lakes. Together with the trends identified in Sections~\ref{sec:substrate} - \ref{sec:exp_setup}, these results point toward plasma tholins with high dispersive surface energies as the most ideal analogs for Titan's atmospheric hazes in terms of their surface properties. Note that even though our samples have highly variable polar components, we cannot use existing Titan observations to distinguish which sample/experimental setup better represents the actual hazes on Titan. As shown in Equation~\ref{eqn6}, only the dispersive component goes into the cloud nucleation calculation, as the observed methane and ethane clouds both have zero polar components. Future work investigating cloud nucleation between tholin and polar condensates observed in Titan's stratosphere, such as HCN and HC$_3$N ice \citep{deKok+2014, Anderson+2018, Hanson+2023}, may leverage differences in polar components across experimental setups to further constrain which laboratory conditions best replicate Titan's haze properties.

\section{Conclusions}
This comparative study demonstrates how surface energy measurements of Titan haze analogs depend on substrate choice, air exposure history, gas mixture compositions, and experimental setup. Our main conclusions are summarized as follows:
\begin{itemize}
\item If the tholin samples have film thicknesses over 50--100~nm, the choice of substrate minimally affects the derived surface energy. Thus, any common substrates could be used for surface energy characterization and would yield similar results. On the contrary, the samples produced in this study with UV irradiation have very low production rates compared to the cold plasma setups and yield films with sub-nm to nm thicknesses. Therefore, their resulting surface energy is strongly affected by the substrate. While these samples qualitatively exhibit surface energies that are statistically distinguishable from their substrates, we cannot determine their true intrinsic surface energy values unless the sample films exceed the 50--100~nm threshold.
\item Although the total surface energy changes only modestly, the dispersive and polar components change substantially after exposing a tholin sample to ambient air, indicating surface chemistry alteration due to oxygen and water vapor exposure. Therefore, it is best to perform surface characterization of tholin samples under vacuum or an inert environment (dry N$_2$ or Ar) to understand Titan-relevant processes.
\item Methane concentration between 1--10\% CH$_4$ in N$_2$ does not significantly influence the total surface energy of tholin samples, demonstrated by the two sets of samples produced with cold plasma discharge. Thus, the variations in methane concentration across Titan's atmosphere are unlikely to substantially modify the overall cohesion of haze particles. However, there are changes in the composing components of surface energy, suggesting some chemical changes are expected for haze particles that form in Titan's atmosphere at different altitudes. In addition, the derived surface energy and its composing components have non-linear responses to the initial methane concentration. The highest/lowest surface energy and its components are not the end-member gas mixtures we used, highlighting the complexity of N$_2$/CH$_4$ plasma chemistry.
\item Comparisons between tholin samples produced with different setups highlight important chemistry both in the laboratory and on Titan. The consistent high dispersive component across all tholin samples means that Titan's actual hazes likely exhibit similarly high dispersive components. The significant differences in polar components across experimental setups likely reflect differences in the production efficiency of polar-bonded molecules by the varying laboratory energy sources and gas exposure timescales.
\item The consistently high dispersive components of our tholin samples indicate that Titan aerosols should be efficient cloud condensation nuclei for hydrocarbon clouds and would likely sink into Titan's hydrocarbon lakes. The observed ethane ice clouds favor cold plasma tholins over FUV tholins as surface-property analogs of Titan's haze, consistent with independent comparisons based on chemical composition \citep{coll2013can}. However, because the UV tholin films in this study are below the substrate-independence threshold, their intrinsic surface energies remain undetermined; this comparison should therefore be revisited once sufficiently thick UV films can be produced.
\end{itemize}

Future work could focus on producing UV tholins thick enough to directly probe their true intrinsic surface energy, whose values reported here are probably affected by the substrate underneath. Future comparative studies could also expand the comparison to include samples produced using other energy sources, such as radio-frequency plasma (i.e., \citealt{Szopa2006-iu, Sekine+2008, Hirai2023, Husic+2026}) and shorter UV photons that can directly dissociate the triple-bonded nitrogen. It would also be interesting to quantify the time dependence of air exposure to better constrain how long samples may be exposed before permanent alteration is made, as it is technically challenging to perform many common laboratory techniques under dry, O$_2$-free conditions. Finally, future studies should aim to compare the surface energies of these tholin films with those of tholin powders (e.g., the PAMPRE setup; \citealt{Szopa2006-iu}). In addition, developing methods to measure surface energy at Titan-relevant temperatures, such as the recent work of \citet{2026Icar..45717154H}, would yield invaluable insights into the behavior of tholins in Titan's cold environment. Finally, the published surface-energy components of additional Titan condensates \citep{Yu_2023-zd} can be combined with the tholin measurements reported here to extend the present CCN and wetting analysis to C$_6$H$_6$, HCN, HC$_3$N, and other potential cloud-forming species.

\section{Acknowledgements}

E.~C.~Austin, X. Yu, E. White, C. He, C. Pesciotta, E. Sciamma-O'Brien, J.~A. Sebree, J.~R. Montes-Bojorquez, A. Husi\'c, C.~R.~Bond, S.~M. H\"orst, F. Salama, and P. McGuiggan are supported by the NASA Cassini Data Analysis Program Grant 80NSSC24K0203. X. Yu, J.~R. Montes-Bojorquez, S.~M. H\"orst, and P. McGuiggan are also supported by the NASA Cassini Data Analysis Program Grant 80NSSC21K0528. X. Yu is also supported by the NASA Planetary Science Early Career Award 80NSSC23K1108, the NASA Solar System Workings Grant 80NSSC24K0888, the NASA Planetary Data Archiving, Restoration and Tools Grant 80NSSC25M7002, and the Heising-Simons Foundation grant 2023-3936. E. Sciamma-O'Brien and F. Salama are also supported by the NASA PSD Cold Solar System Objects Internal Scientist Funding Model project. We thank the anonymous reviewers for their insightful comments and questions, which significantly strengthened the discussion in this manuscript. In particular, we thank them for prompting a deeper exploration of the structural implications of our measurements, suggesting key connections between the dispersive surface energy component and molecular growth efficiency, and directing us to important literature on surface oxidation of tholins.

\bibliographystyle{aasjournal}
\bibliography{Austin_et_al.bib}
\end{CJK*}
\end{document}